\documentclass[aps,showpacs,twocolumn]{revtex4-1}
\usepackage{epsfig}
\usepackage{graphicx,color,dcolumn}
\usepackage{epstopdf}
\usepackage{amsmath,amssymb}
\usepackage{multirow}
\usepackage{diagbox}
\usepackage{changes}
\usepackage{booktabs}
\usepackage{threeparttable}
\usepackage{subfigure}
\usepackage{hyperref}

\newcommand{\ty}{Yue Tan}
\newcommand{\lxj}{Xuejie Liu}
\newcommand{\cxy}{Xiaoyun chen}
\newcommand{\hhx}{Hongxia Huang}
\newcommand{\jlp}{Jialun Ping}
\newcommand{\yc}{Yancheng}
\newcommand{\nj}{Nanjing}
\newcommand{\jl}{Jinling}

\newcommand{\da}{diquark-antidiquark}

\newcommand{\QCD}{QCD}
\newcommand{\cc}{charmonium}
\newcommand{\bb}{bottomium}
\newcommand{\quarkonium}{quarkonium}
\newcommand{\tetraquark}{tetraquark}

\begin{document}
\title{ The newly observed $\Upsilon(10753)$ as a tetraquark  state  in a chiral quark model with scalar nonet exchange}
\author{\ty}
\email[E-mail: ]{tanyue@ycit.edu.cn}
\affiliation{Department of Physics, \yc ~Institute of Technology, \yc~ 224000, P. R. China}

\author{\lxj}
\email[E-mail: ]{1830592517@qq.com}
\affiliation{School of Physics, Southeast University, \nj~ 210094, P. R. China}

\author{\cxy}
\email[E-mail: ]{xychen@jit.edu.cn}
\affiliation{College of Science, \jl ~Institute of Technology, \nj~ 211169, P. R. China}

\author{\hhx}
\email[E-mail: ]{\hhx @njnu.edu.cn (Corresponding author)}
\affiliation{
Department of Physics, \nj~ Normal University, \nj~ 210023, P.R. China
}
\author{\jlp}
\email[E-mail: ]{jlping@njnu.edu.cn (Corresponding author)}
\affiliation{
Department of Physics, \nj~ Normal University, \nj~ 210023, P.R. China
}
\date{\today}

\begin{abstract}
Recently, Belle\uppercase\expandafter{\romannumeral2} Collaboration firstly reported a new resonance $\omega\chi_{bJ}$ in the processes of $e^{+}e^{-} \rightarrow \omega\chi_{bJ}$ at center-of-mass energies $\sqrt{s}=10.745$ GeV. Given the Born cross section similar with the previously reported $\Upsilon(10753)$, the new  resonance $\omega\chi_{bJ}$ may have be $\Upsilon(10753)$. From the perspectives of traditional $\Upsilon(nS)$ meson and exotic tetraquark $b\bar{q}q\bar{b}$ state with the $J^{PC}$ = $1^{--}$, we tentatively investigate the $\Upsilon(10753)$ by solving Schr\"odinger equation in the framework of the chiral quark model. Numerical results for the meson show that the mass of $\Upsilon(5S)$ is up to 10.86 GeV and unsuitable for the candidate of $\Upsilon(10753)$. On the other hand, not only the two kinds of molecular structure ($b\bar{b}$-$q\bar{q}$, $b\bar{q}$-$q\bar{b}$)  but also the \da~structure ($\bar{b}\bar{q}$-$qb$) are considered into our tetraquark calculation with the help of  Gaussian expansion method. When a fully-channel coupling is performed using the real-scaling method  (stabilization method), we get a stable resonance $^1R_{1}(10770)$ ( $^{2S+1}R_J$ ) with a great component of the diquark-antidiquark state. Combined with the mass and width of $^1R_{1}(10770)$, it may be a good candidate for experimental $\Upsilon(10753)$. Besides, several resonance states ranging from 10.82 GeV to 10.96 GeV  are obtained, which are expected to be further verified in future experiments.
\end{abstract}

\maketitle

\section{Introduction} \label{introduction}
Since the $X(3872)$ reported by Belle Collaboration in 2003\cite{Choi:2003}, the different experiments, such as the LHCb, ATLAS, CMS, BESIII, Belle, BABAR, CDF and D$0$, have observed more than 20 exotic states candidates containing $c\bar{c}$ and $b\bar{b}$ quark pairs. These exotic states are beyond the normal meson in the framework of quantum chromo-dynamics (\QCD)  and quark model , and they may contain more information about the low-energy \QCD ~than that of conventional hadrons.

Very recently, the Belle\uppercase\expandafter{\romannumeral2} Collaboration \cite{Belle-II:2022xdi} observed a new structure in the $\chi_{b1}\omega$ invariant mass spectrum with 11~$\sigma$ significance. As Belle\uppercase\expandafter{\romannumeral2} Collaboration reported,  by combining Belle\uppercase\expandafter{\romannumeral2} data with Belle results at $\sqrt{s}$ = 10.867 GeV, they find energy dependencies of the Born cross sections for $e^{+}e^{-}\rightarrow  \omega\chi_{b1,b2}(1P)$ to be consistent
with the shape of the $\Upsilon(10753)$ state. The newly observed structure may be $\Upsilon(10753)$, and we denote the new structure  as $\Upsilon(10753)$ in this paper. What's more, its decay mode $\chi_{bJ}\omega$ indicates that $\Upsilon(10753)$ may be consist of $b\bar{b}$ quark pair and $q\bar{q}$ quark pair. Because it's mass is above the threshold of $\chi_{bJ}\omega$, the possible candidates of $\Upsilon(10753)$ are colorful sub-clusters $bq$-$\bar{q}\bar{b}$ or radial excited $b\bar{b}$. The Belle Collaboration also suggested that the quantum numbers of the state are $J^{P}=1^{-}$.
\begin{figure}[h]
 \vspace{-1.3cm}
\setlength {\abovecaptionskip} {-1.4cm}
  \centering
  \includegraphics[width=9cm,height=13cm]{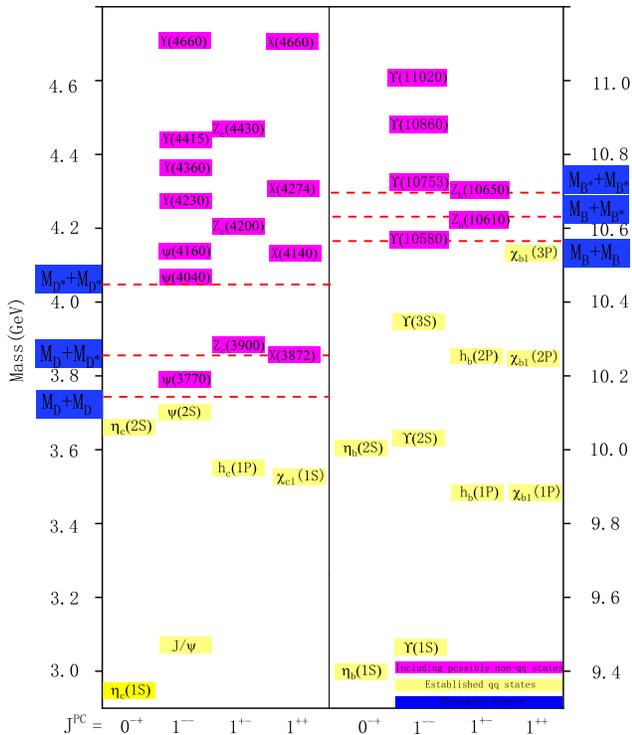}
  \caption{Energy spectrum of $c\bar{c}$ and $b\bar{b}$ system.}
  \label{XYZ}
\end{figure}
\begin{eqnarray*}
M_{\Upsilon(10753)}&=&10753\pm{6.0} ~ \mbox{MeV}  \\ \nonumber
\Gamma_{\Upsilon(10753)}&=& 36_{-12}^{+18} ~ \mbox{MeV}
\end{eqnarray*}

As a heavy \quarkonium ~system, there is a big difference between \cc~ system and \bb~ system in quantitative terms of exotic states.  There are only two certain  exotic candidates $Z_{b}(10610)$ and $Z_b(10650)$ in \bb~system previously reported by the Belle Collaboration\cite{Belle:2011aa,Belle:2015upu}. The leading reason is that the lowest threshold, $DD$, of \cc~system is less than almost half of $c\bar{c}$ states while the mass of he lowest threshold, $BB$, of \bb~system is only below several fourth radial excited $b\bar{b}$ particles such as $\Upsilon(5S)$, two charged $Z_b^+$ and so on Seen in Fig.\ref{XYZ} (data taken from PDG), which means it's difficult for most of particles in the  $b\bar{b}$ system to decay into two B mesons. Thus, the new structure in the $\chi_{b1}\omega$ invariant mass spectrum not only makes the \bb~like family abundant, but also raises our interest in negative parity of $b\bar{b}q\bar{q}$ system. Because the Belle Collaboration has reported $\Upsilon(10753)$ in the process of $e^{+}e^{-} \rightarrow \Upsilon(nS) \pi^{+}\pi^{-}$, there are a lot of researches about negative parity of $b\bar{b}q\bar{q}$ before, which can be interpreted  a traditional bottomonium, tetraquark state or hybrid. In Ref\cite{Chen:2019uzm}, with incorporated the spin-dependent interactions into relativistic flux tube model, the authors interpreted $\Upsilon(10753)$ as a $3^3D_1$ $b\bar{b}$ state. Similarly, the $\Upsilon(10753)$ may be interpreted as  a S-D mixing state dominated by D-wave component in Ref.\cite{Li:2019qsg}. However, the author also poses a  question that what's the mechanism cause such bigger mixing between  $\Upsilon(5S)$ and $\Upsilon(4D)$. In fact, the mass of b quark far outweighing  the other quarks  may lead to very smaller Tensor force and spin-orbit matrix element, which may decrease the coupling effect between S- and D-wave component. Thus some people hold the view that the diquark-antidiquark tetraquark state may be responsible for the existence of  $\Upsilon(10753)$. In Ref.\cite{Wang:2019veq}, Wang \emph{et.al.} assigned $\Upsilon(10753)$ as the diquark-antidiquark type vector hidden-bottom tetraquark state  by QCD sum rules.  Wagner \emph{et.al.}\cite{Bicudo:2020qhp} investigated  $\Upsilon(10753)$ by considering quarkonium and meson-meson components of $\Upsilon(nS)$ including $\Upsilon(10860)$, and confirmed that $\Upsilon(10753)$ may have 76$\%$ meson-meson content. In Ref.\cite{Ali:2019okl}, the author treated $\Upsilon(10753)$ as a linear combination of the diquark-antidiquark and $b\bar{b}$ components due to the mixing via gluonic exchanges.

If the $\Upsilon(10753)$ can be interpreted as compact $bq$-$\bar{q}\bar{b}$ tetraquark, the interaction between two subcluster may be mainly influenced by iso-scalar $q$-$\bar{q}$ pair. However, the $\rho$-$\omega$ (iso-vector $q\bar{q}$ - iso-scalar $q\bar{q}$) mass-reverse  is the long-standing problem with the quark models\cite{Vijande:2004he,Yang:2009zzp} which are lacking of  an isospin-dependent mechanism in the light quark sector. In Ref.\cite{Vijande:2009pu}, the author adds extra scalar meson exchanges such as $a_0$, $f_0$ and $\kappa$ into Gold-stone boson exchange. With this treatment, they successfully obtained $\rho$ and $\omega$ that are consistent with the experiment. However, for keeping in the most economical way, they take several parameters to replace these extra scalar meson exchange. In this writing, instead of adding extra parameters, we investigate meson spectrum in framework of chiral quark model with  scalar nonet exchange based on SU(3) symmetry and hope that this treatment can be better described iso-scalar $b\bar{q}q\bar{b}$ system. Considering the significance of the basis expansion of the trial wave function in the Rayleigh-Ritz variational method, we adopt Gaussian expansion method (GEM) in which each relative motion in the system is expanded in terms of Gaussian basis functions whose sizes are taken in geometric progression. In the tetraquark calculation, we take the mixing effect of molecular structure  and \da~structure into account. Actually, if considering sufficient excited bases including  radial excited bases and spatial excited bases, no matter which structure can describe hadronic systems well. However, it's difficult for us to put too many excited bases into our calculation. An economic way is to combine different structures, and keep the sub-clusters in the low-lying states to do the calculation. Of course, there is a problem that combining different structures could cause over-calculation. In this approach, one should remove strange bases in the Hamiltonian matrix which eigenvalue is 0, and reintegrate the rest bases into new  Hamiltonian matrix. Finally, with the help of ``real scaling method", we can confirm  bound states and  genuine resonances and obtain their decay widths of resonances.
%%%

The paper is organized as follow. After introduction, details of ChQM and GEM are introduced in Section II. In Section III, we present the numerical results and a method of finding and calculating the decay width of the genuine resonance state.  The last section is devoted to the summary.

\section{Chiral quark model, wave function of $b\bar{q}q\bar{b}$ system} \label{wavefunction and chiral quark model}
\subsection{Chiral quark model}
The chiral quark model has been applied successfully in describing the hadron spectra and hadron-hadron interactions.
The details of the model can be found in Refs. \cite{Vijande:2004he,Hu:2021nvs,Tan:2020ldi}. In this writing, we introduce a chiral quark model with scalar nonet exchange.
The Hamiltonian of the chiral quark model is given as follows,
\begin{eqnarray}
H &=& \sum_{i=1}^nm_i+\sum_{i=1}^n(\frac{p_i^2}{2m_i}-T_{CM}) +  \sum_{i<j=1}^n [ V_{con}(r_{ij}) \nonumber\\
&+&V_{oge}(r_{ij}) \sum_{\chi=\pi,\eta,K} V_{\chi}(r_{ij}) +\sum_{s=a_0,f_0,\kappa} V_{s}(r_{ij}) ],
\end{eqnarray}
where $m_i$ is the constituent mass of $i$-th quark (antiquark), and $\mu$ is the reduced masse of two interacting quarks or quark-clusters. $T_{CM}$ is the kinetic energy of the center-of mass motion.  For two quark structure, the kinetic energy term $\sum_{i=1}^n(\frac{p_i^2}{2m_i}-T_{CM})$ will degenerate into $\frac{p_{12}}{2\mu_{12}}$,  while kinetic energy term of tetraquark system can be written as $\frac{p_{12}}{2\mu_{12}}$+$\frac{p_{34}}{2\mu_{34}}$+$\frac{p_{12,34}}{2\mu_{12,34}}$.
\begin{eqnarray}
\mu_{ij}&=&\frac{m_{{i}}  m_{{j}}}{m_{{i}} + m_{{j}}}, ~~ij=12,34   \\
\mu_{1234}&=&\frac{(m_1+m_2)(m_3+m_4)}{m_1+m_2+m_3+m_4},\\
p_{ij}&=&\frac{m_jp_i-m_ip_j}{m_i+m_j},  \\
p_{1234}&=&\frac{(m_3+m_4)p_{12}-(m_1+m_2)p_{34}}{m_1+m_2+m_3+m_4}.
\end{eqnarray}

The different terms of the potential contain central, spin-orbit contributions and tensor force. However, it's difficult for us to deal with spin-orbit and tensor force contributions when  it comes to the computation  of four quarks. Thus, Spin-orbit coupling, tensor force and central force would contribute to our two-quark calculation while only central force would contribute to our four quark calculation.

 $V_{con}(r_{ij})$ is the confining potential, mimics the ``confinement" property of QCD. The $V_{con}(r_{ij})$ term includes central force $V_{con}^{C}(r_{ij})$ and spin-orbit force $V_{con}^{SO}(r_{ij})$.

\begin{align}
\begin{split}
 \left \{
\begin{array}{ll}
    V_{con}^C(r_{ij}) &= ( -a_{c} r_{ij}^{2}-\Delta) \boldsymbol{\lambda}_i^c \cdot \boldsymbol{\lambda}_j^c\\
\\
    V_{con}^{SO}(r_{ij}) &= -\boldsymbol{\lambda}_i^c \cdot \boldsymbol{\lambda}_j^c \frac{a_c}{4m_i^2m_j^2}[ ((m_i^2+m_j^2)(1-2a_s) \\
    &+ 4m_im_j(1-a_s))(\vec{S}_{+}\cdot \vec{L})+\\
    &((m_j^2-m_i^2)(1-2a_s))(\vec{S}_{-}\cdot \vec{L}) ]\\
  %  V_{con}^{SO}(r_{ij}) = -\boldsymbol{\lambda}_i^c \cdot \boldsymbol{\lambda}_j^c \frac{a_c}{4m_i^2m_j^2}[ ((m_i^2+m_j^2)(1-2a_s)+4m_im_j(1-a_s))(\vector{S}_{+}\cdot \vector{L}) ]\\
\end{array}
\right.
\end{split}
\end{align}

The second potential $V_{oge}(r_{ij})$ is one-gluon exchange interaction reflecting the ``asymptotic freedom" property of QCD. The $V_{oge}(r_{ij})$ term contains central force $V_{oge}^{C}(r_{ij})$, spin-orbit force $V_{oge}^{SO}(r_{ij})$ and tensor force $V_{oge}^{T}(r_{ij})$.

\begin{align}
\begin{split}
 \left \{
\begin{array}{ll}
    V_{oge}^C(r_{ij}) &=\frac{\alpha_s}{4} \boldsymbol{\lambda}_i^c \cdot \boldsymbol{\lambda}_{j}^c
\left[\frac{1}{r_{ij}}-\frac{1}{6m_im_jr_0^2}\boldsymbol{\sigma}_i\cdot
\boldsymbol{\sigma}_j \frac{e^{-r_{ij}}/r_0(\mu_{ij})}{r_{ij}}\right]   \\
%&   \delta{(\boldsymbol{r}_{ij})}=\frac{e^{-r_{ij}/r_0(\mu_{ij})}}{4\pi r_{ij}r_0^2(\mu_{ij})}, \nonumber\\
    \\
    V_{oge}^{SO}(r_{ij}) &= -\frac{1}{16} \frac{\alpha_s\boldsymbol{\lambda}_i^c \cdot \boldsymbol{\lambda}_j^c}{4m_i^2m_j^2}[\frac{1}{r_{ij}^3}-\frac{e^{-r_{ij}/r_g(\mu)}}{r_{ij}^3}(1+\frac{r_{ij}}{r_g(\mu)})] \\
    &[ (m_i^2+m_j^2+4m_im_j)(\vec{S}_{+}\cdot \vec{L})\\
    &+(m_j^2-m_i^2)(\vec{S}_{-}\cdot \vec{L}) ]\\
V_{oge}^{T}(r_{ij}) &= -\frac{1}{16} \frac{\alpha_s\boldsymbol{\lambda}_i^c \cdot \boldsymbol{\lambda}_j^c}{4m_i^2m_j^2}[\frac{1}{r_{ij}^3}-\frac{e^{-r_{ij}/r_g(\mu)}}{r_{ij}}(\frac{1}{r_{ij}^2}\\
&+\frac{1}{3r^2_g(\mu)} +\frac{1}{r_{ij}r_g(\mu)})]S_{ij} \\
  %  V_{con}^{SO}(r_{ij}) = -\boldsymbol{\lambda}_i^c \cdot \boldsymbol{\lambda}_j^c \frac{a_c}{4m_i^2m_j^2}[ ((m_i^2+m_j^2)(1-2a_s)+4m_im_j(1-a_s))(\vector{S}_{+}\cdot \vector{L}) ]\\
\end{array}
\right.
\end{split}
\end{align}

$\boldsymbol{\sigma}$ are the $SU(2)$ Pauli matrices; $\boldsymbol{\lambda}_{c}$ are $SU(3)$ color Gell-Mann matrices,
$r_{0}(\mu_{ij})=\frac{r_0}{\mu_{ij}}$ and $\alpha_{s}$ is an effective scale-dependent running coupling,
\begin{equation}
 \alpha_s(\mu_{ij})=\frac{\alpha_0}{\ln\left[(\mu_{ij}^2+\mu_0^2)/\Lambda_0^2\right]}.
\end{equation}
The third potential $V_{\chi, s}$ is Goldstone boson exchange, coming from ``chiral symmetry spontaneous breaking" of QCD
in the low-energy region. There are two different contributions including pseudoscalar meson exchange $V_{\chi=\pi,\eta,K}(r_{ij})$ and scalar meson exchange $V_{s=a_0,f_0,\kappa}(r_{ij})$ to the Goldstone boson exchange. The pseudoscalar meson exchange $V_{\chi=\pi,\eta,K}(r_{ij})$  contains central force and tensor force while the scalar meson exchange $V_{s=a_0,f_0,\kappa}(r_{ij})$ consist of central force contribution and spin-orbit contribution.

\begin{eqnarray}
V_{\chi}(r_{ij}) & = & v_{\pi}({{\bf r}_{ij}})\sum_{a=1}^{3}
	\lambda_i^a \lambda_j^a+v_{K}({{\bf r}_{ij}})\sum_{a=4}^{7}
	\lambda_i^a \lambda_j^a+v_{\eta}({{\bf r}_{ij}}) \nonumber \\
 & &  [\cos\theta_{P}(\lambda_i^8 \lambda_j^8)-\sin\theta_{P}(\lambda_i^0 \lambda_j^0)] ,  \\
v_{\chi=\pi,K,\eta} & = & \frac{g^2_{ch}}{4\pi}\frac{m^2_\chi}{12m_im_j}
	\frac{\Lambda^2_\chi}{\Lambda^2_\chi-m^2_\chi}m_\chi \nonumber \\
& &	\left[ Y(m_{\chi}r_{ij})-\frac{\Lambda^3_\chi}{m^3_\chi}Y (\Lambda_{\chi}r_{ij}) \right]
	(\boldsymbol{\sigma}_i \cdot \boldsymbol{\sigma}_j),  \nonumber \\
V_{S}(r_{ij}) & = & v_{\sigma}({{\bf r}_{ij}})
	\lambda_i^0 \lambda_j^0+v_{a_0}({{\bf r}_{ij}})\sum_{a=1}^{3}
	\lambda_i^a \lambda_j^a   \nonumber \\
& & +v_{\kappa}({{\bf r}_{ij}})\sum_{a=4}^{7}
	\lambda_i^a \lambda_j^a+v_{f_0}({{\bf r}_{ij}})
    \lambda_i^8 \lambda_j^8
 \end{eqnarray}

 \begin{align}
\begin{split}
 \left \{
\begin{array}{ll}
v_{s=\sigma,a_0,f_0,\kappa}^C & =  -\frac{g^2_{ch}}{4\pi} \frac{\Lambda^2_s}{\Lambda^2_s-m^2_s}m_s
	\left[ Y(m_{s}r_{ij})-\frac{\Lambda_s}{m_s}Y(\Lambda_{s}r_{ij})\right],   \\
v_{s=\sigma,a_0,f_0,\kappa}^{SO} & =  -\frac{g^2_{ch}}{4\pi} \frac{\Lambda^2_s}{\Lambda^2_s-m^2_s}\frac{m_s^3}{2m_im_j}
	[ G(m_{s}r_{ij}) \\
&-\frac{\Lambda_s^3}{m_s^3}G(\Lambda_{s}r_{ij})]\vec{L}\cdot \vec{S},
\end{array}
\right.
\end{split}
\end{align}
\begin{align}
\begin{split}
 \left \{
\begin{array}{ll}
v_{s=\pi,K,\eta}^C & =  \frac{g^2_{ch}}{4\pi} \frac{m_{s}^2}{\Lambda_s^2-m_s^2} \frac{\Lambda^2_s}{\Lambda^2_s-m^2_s}m_s[ Y(m_{s}r_{ij})-\frac{\Lambda_s^3}{m_s^3}\\
&Y(\Lambda_{s}r_{ij})](\vec{\sigma}_i\cdot\vec{\sigma}_j),   \\
v_{s=\pi,K,\eta}^T & =  \frac{g^2_{ch}}{4\pi} \frac{m_{s}^2}{\Lambda_s^2-m_s^2} \frac{\Lambda^2_s}{\Lambda^2_s-m^2_s}m_s
	[ H(m_{s}r_{ij})-\frac{\Lambda_s^3}{m_s^3}\\
&H(\Lambda_{s}r_{ij})]S_{ij},   \\
\end{array}
\right.
\end{split}
\end{align}
$\boldsymbol{\lambda}$ are $SU(3)$ flavor Gell-Mann matrices, $m_{\chi(s)}$ is the masses of Goldstone bosons, $\Lambda_{\chi(s)}$ is the cut-offs, $g^2_{ch}/4\pi$ is the Goldstone-quark coupling constant.  The quark tensor operator is determined by $S_{ij}=3(\vec{\sigma}_i\cdot\hat{r}_{ij})(\vec{\sigma}_j\cdot\hat{r}_{ij})-(\vec{\sigma}_i\cdot\vec{\sigma}_j)$. Finally, $Y(x)$ is the standard Yukawa function defined by $Y(x)=e^{-x}/x$, $G(x)=(1+1/x)Y(x)/x$ and $H(x)=(1+3/x+3/x^2)Y(x)/x$.

All the parameters are determined by fitting the meson spectrum, from light mesons ($\pi, \eta$) to heavy mesons ($\chi_{bJ}$), taking into account only a
quark-antiquark component. They are shown in Table~\ref{modelparameters}. The calculated masses of the mesons involved
in the present work are shown in Table~\ref{mesonmass}. %Because the spin-orbit interaction is not considered here,
%we obtained a degenerate eigen-energy for the three $P$-wave states, $^3P_J,J=0,1,2$.

\begin{table}[t]
\begin{center}
\caption{Quark model parameters ($m_{\pi}=0.7$ fm, $m_{\sigma}=3.42$ fm, $m_{\eta}=2.77$ fm).\label{modelparameters}}
\begin{tabular}{cccc}
\hline\hline\noalign{\smallskip}
Quark masses   &$m_u=m_d$(MeV)     &490  \\
               &$m_{b}$(MeV)         &4978 \\
\hline
Goldstone bosons
                   &$\Lambda_{\pi}(fm^{-1})$     &3.5  \\
                   &$\Lambda_{\eta}(fm^{-1})$    &2.2  \\
                   &$\Lambda_{a0}(fm^{-1})$      &2.5  \\
                   &$\Lambda_{f0}(fm^{-1})$      &1.2  \\
                   &$g_{ch}^2/(4\pi)$                &0.54  \\
                   &$\theta_p(^\circ)$                &-15 \\
\hline
Confinement             &$a_{c}$(MeV)     &98 \\
                   &$\Delta$(MeV)       &-18.1 \\
\hline
OGE                & $\alpha_{qq}$        &1.34 \\
                   & $\alpha_{qb}$        &0.75 \\
                   & $\alpha_{bb}$        &0.43 \\
                   &$\hat{r}_0$(MeV)    &81.0 \\
                   &$\hat{r}_g$(MeV)    &100.6 \\
                   & $a_{s}$        &0.77 \\
\hline\hline
\end{tabular}
\end{center}
\end{table}

\begin{table}[]
\caption{ \label{mesonmass}  Numerical results for the meson spectrum (in MeV) for the ChQM1( with a  harmonic form confinement ), the ChQM2( with a  color screening form confinement ) and Relative quark model. (unit: MeV).\label{mesons}}
\begin{tabular}{cccccc}
\hline\hline\noalign{\smallskip}
    Meson         &     ChQM1& ChQM2\cite{Vijande:2009pu} &     RM\cite{Deng:2016ktl} & EXP.(PDG) \\ \hline
    $\pi$         &     149  &   139 &    - & 139.57$\pm$0.00035\\
    $\eta$        &     524  &   572 &   - & 547.862$\pm$0.017\\
    $\rho$        &     765  &   775 &   - & 775.26$\pm$0.34\\
    $\omega$      &     780  &   691 &   - & 782.66$\pm$0.13\\
    $B$           &    5273  &  5281 &   - & 5279.66$\pm$0.12\\
    $B^*$         &    5329  &  5321 &   - & 5324.71$\pm$0.21\\
    $B_0$         &    5770  &    -  &   - & -               \\
    $B_1$         &    5813  &    -  &   - & $5725.9^{+2.5}_{-2.7}$\\
    $B_2$         &    5850  &  5790 &   - & 5735.2$\pm$0.7\\
    $\eta_b$      &    9390  &  9454 &   9390 & 9398.7$\pm$2.0\\
    $\Upsilon(1S)$&    9499  &  9505 &   9460 & 9460.3$\pm$0.26\\
    $\Upsilon(2S)$&    9968  & 10013 &   10015 & 10023.6$\pm$0.31\\
    $\Upsilon(3S)$&   10278  & 10335 &   10343 & 10355.2$\pm$0.5\\
    $\Upsilon(4S)$&   10564  & 10577 &   10579 & 10579.4$\pm$1.2\\
    $\Upsilon(3D)$&   10631  & -     &      -  & 10579.4$\pm$1.2\\
    $\Upsilon(5S)$&   10863  & 10770 &   10811 & $10885.2^{+2.6}_{-1.6}$\\
    $h_b$         &    9897  &   -   &   9909 & 9899.3$\pm$0.8\\
    $\chi_{b0}$   &    9865  &  9855 &   9864 & 9859.44$\pm$0.42$\pm$0.31\\
    $\chi_{b1}$   &    9891  &  9875 &   9903 & 9892.78$\pm$0.26$\pm$0.31\\
    $\chi_{b2}$   &    9908  &  9887 &   9921 & 9912.21$\pm$0.26$\pm$0.31\\
\hline\hline
\end{tabular}
\end{table}

\subsection{The wave function of $b\bar{q}q\bar{b}$ system}
\begin{figure*}[htbp]
\centering

\subfigure[]{
\begin{minipage}[t]{0.2\linewidth}
\centering
\includegraphics[width=3.5cm,height=2.8cm]{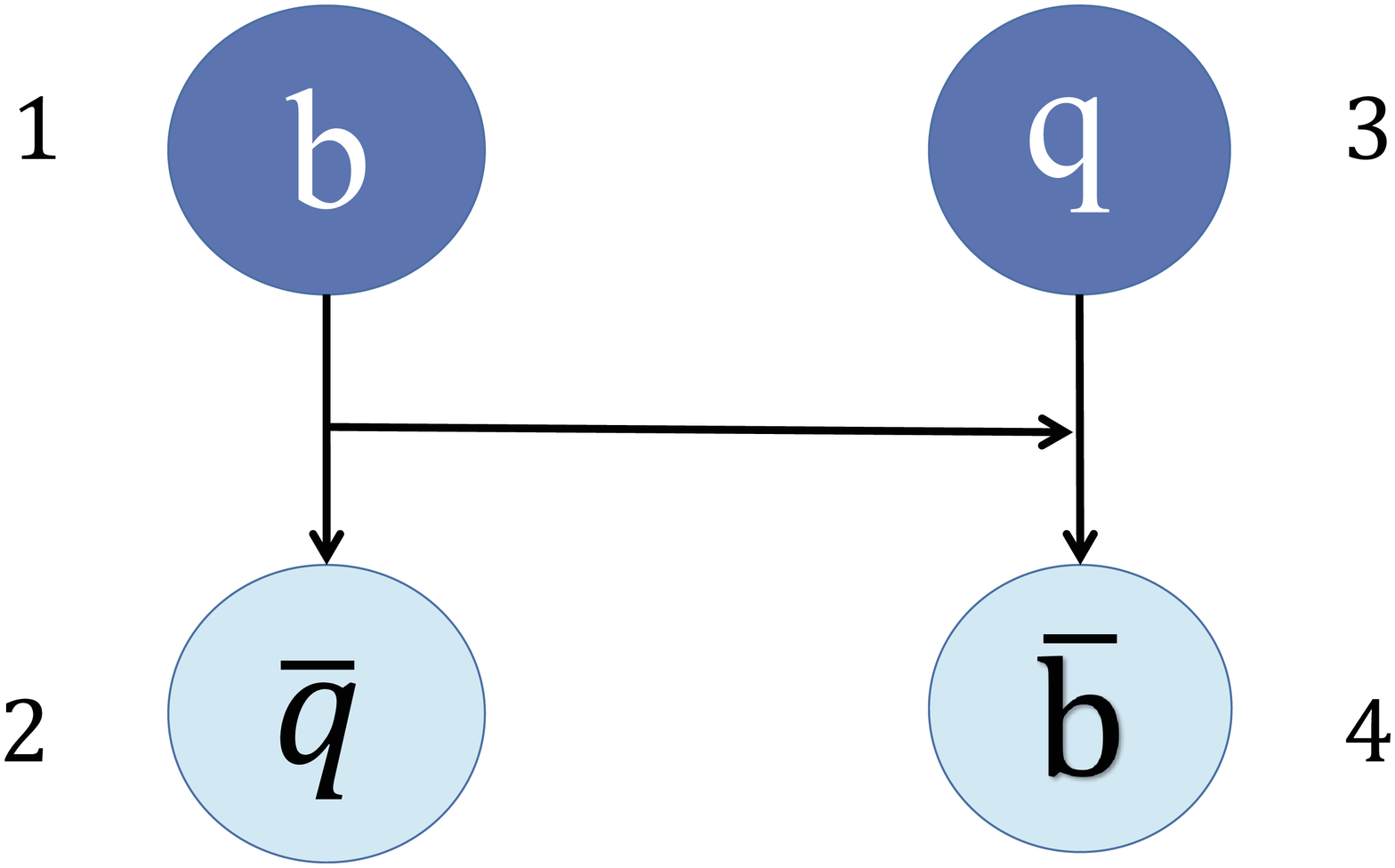}
%\caption{fig1}
\end{minipage}%
}%
\subfigure[]{
\begin{minipage}[t]{0.2\linewidth}
\centering
\includegraphics[width=3.5cm,height=2.8cm]{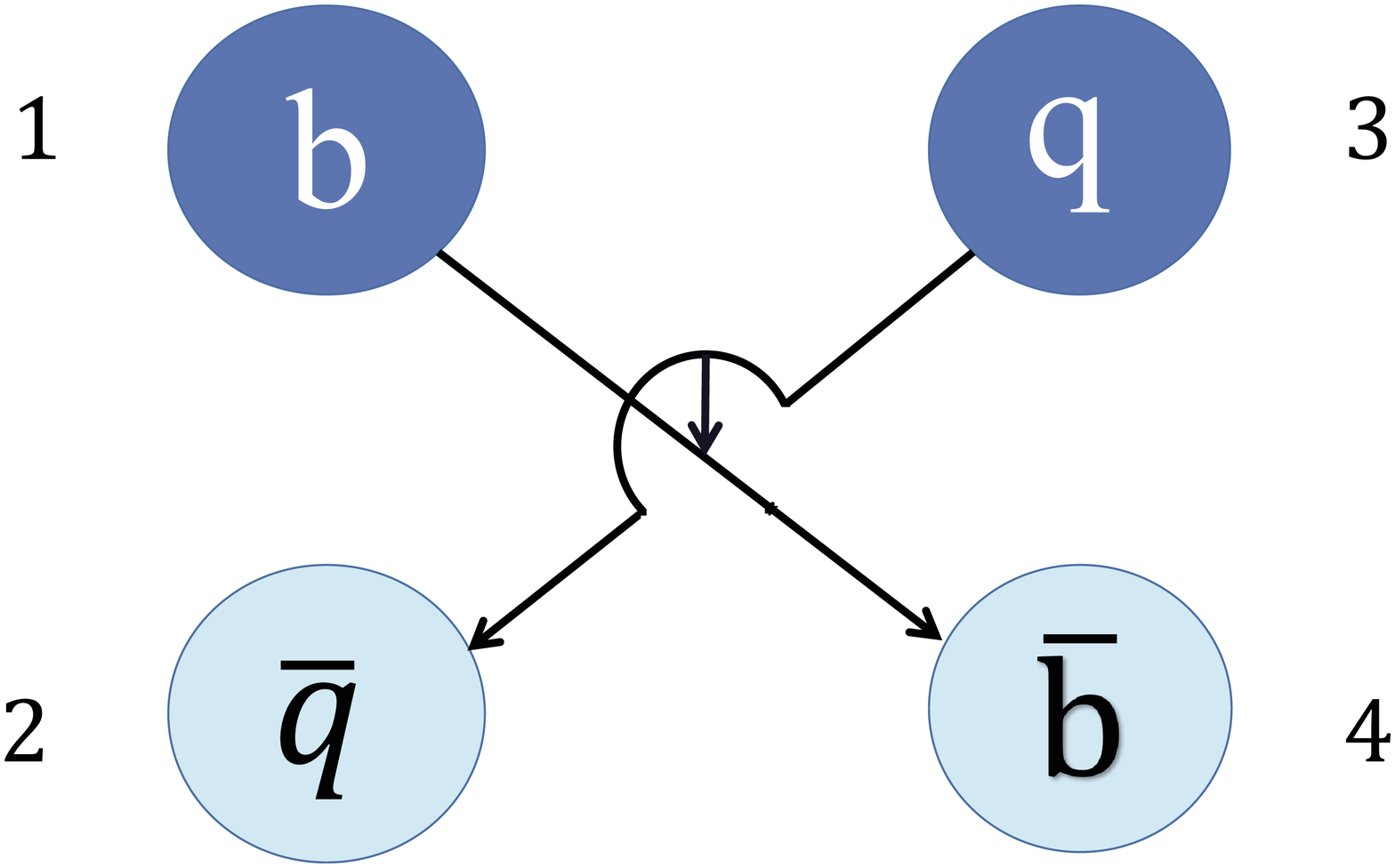}
%\caption{fig2}
\end{minipage}%
}% %这个回车键很重要 \quad也可以
\subfigure[]{
\begin{minipage}[t]{0.1\linewidth}
\centering
\includegraphics[width=3.5cm,height=2.8cm]{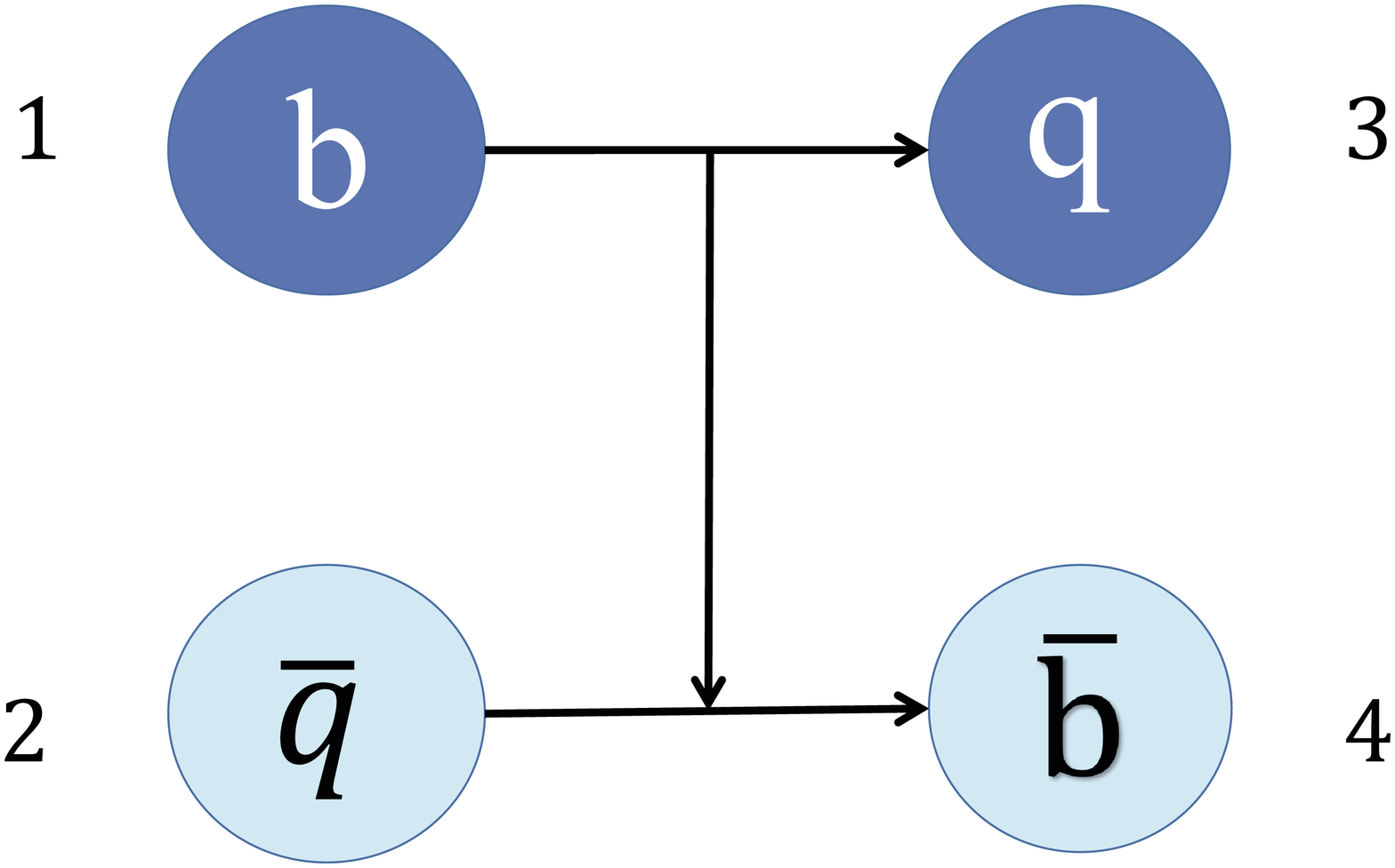}
%\caption{fig2}
\end{minipage}
}%

\centering
\caption{ Three kinds of configuration of $b$, $\bar{b}$, $q$ and $\bar{q}$  system (a) the molecular configuration of
$bq\bar{q}\bar{b}$ system; (b) the molecular configurations of $bb\bar{q}\bar{q}$ system; (c) the \da~ configurations of $bq\bar{b}\bar{q}$ system;}.\label{Fic}
\end{figure*}

There are two physically important structures, meson-meson and diquark-antidiquark in the $b\bar{b}q\bar{q}$ system, which are considered in the present calculation.
The wave functions of every structure all consists of four parts: orbital, spin, flavor and color. The wave function of each part is constructed in two steps, first write down the two-body wave functions, then coupling two sub-clusters wave functions to form the four-body one. To distinguish different structures, we give very quark with specific particle order ($b_1$, $\bar{q}_2$, $q_3$, $\bar{b}_4$). As a consequence, the $b_1$ coupled with $\bar{q}_2$ while the $q_3$ coupled with $\bar{b}_4$ can lead to dimeson $b_1\bar{q}_2$-$q_3\bar{b}_4$  structure. If  we couple the $b_1\bar{b}_3$ into the first sub-cluster and   $q_2\bar{q}_4$ into the second sub-cluster, the $b_1\bar{b}_4$-$q_3\bar{q}_2$-type dimeson states can be obtained. As for diquark-antidiquark structure, it's particle order is $b_1q_3$-$\bar{q}_2\bar{b}_4$. All of the three structures can be found in Fig. \ref{Fic}. In this treatment, its convenient for us to do coupling calculation between different structures. With the certain isospin, three kinds of flavor wave-functions are obtained in Eq.\ref{sec}. Considering that the difference of each structure is particle order, we only present the spatial wave-function, spin wave-function and color wave-function of the $b_1\bar{q}_2$-$q_3\bar{b}_4$ dimeson structure and the other color wave-function of the $b_1q_3$-$q_2b_4$ diquark-antidiquark structure in the next subsection for simplicity. However, we should keep in mind, when we deal with the other structures, the particles order should be adjusted correctly. Because there are no identical particles in the system, the total wave function of the system is the direct product of spatial ($|R_{i}\rangle$), spin ($|S_{j}\rangle$), color ($|C_{k}\rangle$) and flavor ($|F_{n}\rangle$) wave functions with necessary coupling.
%\begin{equation}\label{bohanshu}
%|ijkn\rangle=[|R_{i}\rangle\otimes|S_{j}\rangle]\otimes|C_{k}\rangle\otimes |F_{n}\rangle
%\end{equation}

\subsubsection{flavor wave function\label{sec_flavor}}

We have three flavor wave functions of the system,
\begin{eqnarray}\label{sec}
|F_{1}\rangle&=\frac{1}{\sqrt{2}}( b_1\bar{u}_2u_3\bar{b}_4+b_1\bar{d}_2d_3\bar{b}_4),\\
|F_{2}\rangle&=\frac{1}{\sqrt{2}}( b_1\bar{b}_4u_3\bar{u}_2+b_1\bar{b}_4d_3\bar{d}_2),\\
|F_{3}\rangle&=\frac{1}{\sqrt{2}}( b_1u_3\bar{u}_2\bar{b}_4+b_1d_3\bar{d}_2\bar{b}_4)
\end{eqnarray}
$|F_{1}\rangle, |F_{2}\rangle$ is for meson-meson structure, and $|F_{3}\rangle$ is for diquark-antidiquark structure.

%\begin{figure}[htp]
%  \centering
%  \resizebox{0.50\textwidth}{!}{\includegraphics[width=3cm,height=2.6cm]{M1.eps}~\includegraphics[width=3cm,height=2.6cm]{M2.eps}~\includegraphics[width=3cm,height=2.6cm]{D1.eps}}
%  \caption{\label{MMstructures} Three kinds of configuration of $b$, $\bar{b}$, $q$ and $\bar{q}$  system (a) the molecular configuration of
%$bq\bar{q}\bar{b}$ system; (b) the molecular configurations of $bb\bar{q}\bar{q}$ system; (c) the \da~ configurations of $bq\bar{b}\bar{q}$ system;}
%  \label{1}
%\end{figure}
%
%

\subsubsection{spatial wave function}
The orbital wave function of the four-quark system consists of two sub-cluster orbital wave function and the relative
motion wave function between two subclusters,
\begin{eqnarray}\label{spatialwavefunctions}
|R_{1}\rangle&=&\left[[\Psi_{l_1=1}({\bf r}_{12})\Psi_{l_2=0}({\bf
r}_{34})]_{l_{12}}\Psi_{L_r}({\bf r}_{1234}) \right]_{L}^{M_{L}}, \nonumber\\
|R_{2}\rangle&=&\left[[\Psi_{l_1=0}({\bf r}_{12})\Psi_{l_2=1}({\bf
r}_{34})]_{l_{12}}\Psi_{L_r}({\bf r}_{1234}) \right]_{L}^{M_{L}}.
\end{eqnarray}
where the bracket "[~]" indicates orbital angular momentum coupling, and $L$ is the total orbital angular momentum which
comes from the coupling of $L_r$ ( orbital angular momentum of relative motion ) and $l_{12}$  which is derived by coupling $l_1$ and $l_2$. The $l_1$ ($l_2$) is sub-cluster orbital angular momenta. $|R_{i,i=1,2}\rangle$ are used to represent the orbital wave functions of meson-meson structure,
and $|R_{i=3,4}\rangle$ are used to represent the wave functions of diquark-antidiquark structure.
In GEM \cite{Hiyama:2003cu}, the radial part of the orbital wave function is expanded by a set of Gaussians:
\begin{subequations}
\label{radialpart}
\begin{align}
\Psi(\mathbf{r}) & = \sum_{n=1}^{n_{\rm max}} c_{n}\psi^G_{nlm}(\mathbf{r}),\\
\psi^G_{nlm}(\mathbf{r}) & = N_{nl}r^{l}
e^{-\nu_{n}r^2}Y_{lm}(\hat{\mathbf{r}}),
\end{align}
\end{subequations}
where $N_{nl}$ are normalization constants,
\begin{align}
N_{nl}=\left[\frac{2^{l+2}(2\nu_{n})^{l+\frac{3}{2}}}{\sqrt{\pi}(2l+1)}
\right]^\frac{1}{2}.
\end{align}
$c_n$ are the variational parameters, which are determined dynamically. The Gaussian size parameters are chosen according
to the following geometric progression
\begin{equation}\label{gaussiansize}
\nu_{n}=\frac{1}{r^2_n}, \quad r_n=r_1a^{n-1}, \quad
a=\left(\frac{r_{n_{\rm max}}}{r_1}\right)^{\frac{1}{n_{\rm max}-1}}.
\end{equation}
This procedure enables optimization of the using of Gaussians, as small as possible Gaussians are used.

\subsubsection{spin wave function}
Because of no difference between spin of quark and antiquark, the meson-meson structure has the same spin wave function as
the diquark-antidiquark structure. The spin wave functions of the sub-cluster are shown below.
\begin{align*}
&\chi_{11}^{\sigma}=\alpha\alpha,~~
\chi_{10}^{\sigma}=\frac{1}{\sqrt{2}}(\alpha\beta+\beta\alpha),~~
\chi_{1-1}^{\sigma}=\beta\beta,\nonumber \\
&\chi_{00}^{\sigma}=\frac{1}{\sqrt{2}}(\alpha\beta-\beta\alpha),
\end{align*}
Coupling the spin wave functions of two sub-clusters by Clebsch-Gordan coefficients, total spin wave function can be written below,
%\begin{align*}
%|S_{1}\rangle=\chi_{0}^{\sigma1}&=\chi_{00}^{\sigma}\chi_{00}^{\sigma},\\
%|S_{2}\rangle=\chi_{0}^{\sigma2}&=\sqrt{\frac{1}{3}}(\chi_{11}^{\sigma}
%  \chi_{1-1}^{\sigma}-\chi_{10}^{\sigma}\chi_{10}^{\sigma}+\chi_{1-1}^{\sigma}\chi_{11}^{\sigma}),\\
%|S_{3}\rangle=\chi_{1}^{\sigma1}&=\chi_{00}^{\sigma}\chi_{11}^{\sigma},\\
%|S_{4}\rangle=\chi_{1}^{\sigma2}&=\chi_{11}^{\sigma}\chi_{00}^{\sigma},\\
%|S_{5}\rangle=\chi_{1}^{\sigma3}&=\frac{1}{\sqrt{2}}(\chi_{11}^{\sigma}\chi_{10}^{\sigma}-\chi_{10}^{\sigma}\chi_{11}^{\sigma}),\\
%|S_{6}\rangle=\chi_{2}^{\sigma1}&=\chi_{11}^{\sigma}\chi_{11}^{\sigma}.\\
%\end{align*}

\begin{align}
&|S_{1}\rangle=\chi_{0}^{\sigma1} =  \sqrt{\frac{1}{4}}(\alpha_1\beta_2\alpha_3\beta_4-\alpha_1\beta_2\beta_3\alpha_4-\beta_1\alpha_2\alpha_3\beta_4\\
&+\beta_1\alpha_2\beta_3\alpha_4 ),\\
&|S_{2}\rangle=\chi_{0}^{\sigma2} =  \sqrt{\frac{1}{12}}(2\alpha_1\alpha_2\beta_3\beta_4+2\beta_1\beta_2\alpha_3\alpha_4-\alpha_1\beta_2\alpha_3\beta_4\\
&-\alpha_1\beta_2\beta_3\alpha_4-\beta_1\alpha_2\alpha_3\beta_4-\beta_1\alpha_2\beta_3\alpha_4 ),\\
&|S_{3}\rangle=\chi_{1}^{\sigma3} =  \sqrt{\frac{1}{4}}(\alpha_1\beta_2\alpha_3\beta_4+\alpha_1\beta_2\beta_3\alpha_4-\beta_1\alpha_2\alpha_3\beta_4\\
&-\beta_1\alpha_2\beta_3\alpha_4 ),\\
&|S_{4}\rangle=\chi_{1}^{\sigma4} =  \sqrt{\frac{1}{4}}(\alpha_1\beta_2\alpha_3\beta_4-\alpha_1\beta_2\beta_3\alpha_4+\beta_1\alpha_2\alpha_3\beta_4\\
&-\beta_1\alpha_2\beta_3\alpha_4 ),\\
&|S_{5}\rangle=\chi_{1}^{\sigma5} =  \sqrt{\frac{1}{2}}(\alpha_1\alpha_2\beta_3\beta_4-\beta_1\beta_2\alpha_3\alpha_4 ),\\
&|S_{6}\rangle=\chi_{2}^{\sigma6} =  \sqrt{\frac{1}{6}}(\alpha_1\alpha_2\beta_3\beta_4+\beta_1\beta_2\alpha_3\alpha_4+\alpha_1\beta_2\alpha_3\beta_4\\
&+\alpha_1\beta_2\beta_3\alpha_4+\beta_1\alpha_2\alpha_3\beta_4+\beta_1\alpha_2\beta_3\alpha_4 ).\\
\end{align}
\\
The total spin wave function is denoted by $\chi_{S}^{\sigma i}$, $i$ is the index of the functions, the $S$ is the total spin
of the system. Due to absence of a spin-dependent interaction in the Hamiltonian which can distinguish the third component of the spin quantum
number, we set the third component of the spin to be zero for simplicity.

%%\begin{figure}[htp]
%%\resizebox{0.50\textwidth}{!}{\includegraphics{M1.eps},\includegraphics{M2.eps},\includegraphics{D1.eps}}
%%\caption{\label{MMstructures} Three kinds of configuration of $b$, $\bar{b}$, $q$ and $\bar{q}$  system (a) the molecular configuration of
%%$bq\bar{q}\bar{b}$ system; (b) the molecular configurations of $bb\bar{q}\bar{q}$ system; (c) the \da~ configurations of $bq\bar{b}\bar{q}$ system;}
%%\end{figure}

\subsubsection{color wave function}
The colorless tetraquark system has four color wave functions, two for meson-meson structure, $1\otimes1$ ($C_1$), $8\otimes8$ ($C_2$),
and two for diquark-antidiquark structure, $\bar{3}\otimes 3$ ($C_3$) and $6\otimes \bar{6}$ ($C_4$).

%\begin{align}
%&|C_{1}\rangle =  \sqrt{\frac{1}{9}}( {r_1\bar{r}_2r_3\bar{r}_4+r_1\bar{r}_2g_3\bar{g}_4+r_1\bar{r}_2b_3\bar{b}_4+g_1\bar{g}_2r_3\bar{r}_4  \\
%& +g_1\bar{g}_2g_3\bar{g}_4 +g_1\bar{g}_2b_3\bar{b}_4+b_1\bar{b}_2r_3\bar{r}_4  +b_1\bar{b}_2g_3\bar{g}_4+b_1\bar{b}_2b_3\bar{b}_4) \nonumber\\
%\end{align}
%\begin{eqnarray}
%\begin{aligned}
%|C_{2}\rangle = \sqrt{\frac{1}{72}}(3\bar{b}r\bar{r}b+3\bar{g}r\bar{r}g+3\bar{b}g\bar{g}b+3\bar{g}b\bar{b}g+3\bar{r}g\bar{g}r \nonumber \\
% +3\bar{r}b\bar{b}r+2\bar{r}r\bar{r}r+2\bar{g}g\bar{g}g+2\bar{b}b\bar{b}b-\bar{r}r\bar{g}g \nonumber \\
% -\bar{g}g\bar{r}r-\bar{b}b\bar{g}g-\bar{b}b\bar{r}r-\bar{g}g\bar{b}b-\bar{r}r\bar{b}b). \nonumber\\
%%|C_{3}\rangle &= &
%% \sqrt{\frac{1}{12}}(rg\bar{r}\bar{g}-rg\bar{g}\bar{r}+gr\bar{g}\bar{r}-gr\bar{r}\bar{g}+rb\bar{r}\bar{b} \nonumber \\
%% & & -rb\bar{b}\bar{r}+br\bar{b}\bar{r}-br\bar{r}\bar{b}+gb\bar{g}\bar{b}-gb\bar{b}\bar{g} \nonumber \\
%% & & +bg\bar{b}\bar{g}-bg\bar{g}\bar{b}). \nonumber \\
%%|C_{4}\rangle &= & \sqrt{\frac{1}{24}}(2rr\bar{r}\bar{r}+2gg\bar{g}\bar{g}+2bb\bar{b}\bar{b}
%%    +rg\bar{r}\bar{g}+rg\bar{g}\bar{r} \nonumber \\
%%& & +gr\bar{g}\bar{r}+gr\bar{r}\bar{g}+rb\bar{r}\bar{b}+rb\bar{b}\bar{r}+br\bar{b}\bar{r} \nonumber \\
%%& & +br\bar{r}\bar{b}+gb\bar{g}\bar{b}+gb\bar{b}\bar{g}+bg\bar{b}\bar{g}+bg\bar{g}\bar{b}).\nonumber
%\end{aligned}
%\end{eqnarray}

\begin{align}
&|C_{1}\rangle =  \sqrt{\frac{1}{9}} ( r_1\bar{r}_2r_3\bar{r}_4+r_1\bar{r}_2g_3\bar{g}_4+r_1\bar{r}_2b_3\bar{b}_4+g_1\bar{g}_2r_3\bar{r}_4 \nonumber\\
& +g_1\bar{g}_2g_3\bar{g}_4 +g_1\bar{g}_2b_3\bar{b}_4+b_1\bar{b}_2r_3\bar{r}_4  +b_1\bar{b}_2g_3\bar{g}_4+b_1\bar{b}_2b_3\bar{b}_4)  \nonumber\\
%\end{align*}
%\begin{align*}
&|C_{2}\rangle = \sqrt{\frac{1}{72}}(3r_1\bar{b}_2b_3\bar{r}_4+3r_1\bar{g}_2g_3\bar{r}_4+3g_1\bar{b}_2b_3\bar{g}_4                   \nonumber\\
&+3b_1\bar{g}_2g_3\bar{b}_4+3g_1\bar{r}_2r_3\bar{g}_4+3b_1\bar{r}_2r_3\bar{b}_4+2r_1\bar{r}_2r_3\bar{r}_4 \nonumber\\
&+2g_1\bar{g}_2g_3\bar{g}_4+2b_1\bar{b}_2b_3\bar{b}_4 -r_1\bar{r}_2g_3\bar{g}_4-g_1\bar{g}_2r_3\bar{r}_4\nonumber\\
&-b_1\bar{b}_2g_3\bar{g}_4-b_1\bar{b}_2r_3\bar{r}_4-g_1\bar{g}_2b_3\bar{b}_4-r_1\bar{r}_2b_3\bar{b}_4). \nonumber\\
%\end{align*}
%\begin{align*}
&|C_{3}\rangle = \sqrt{\frac{1}{12}}(r_1g_3\bar{r}_2\bar{g}_4-r_1g_3\bar{g}_2\bar{r}_4+g_1r_3\bar{g}_2\bar{r}_4-g_1r_3\bar{r}_2\bar{g}_4 \nonumber \\
&+r_1b_3\bar{r}_2\bar{b}_4-r_1b_3\bar{b}_2\bar{r}_4+b_1r_3\bar{b}_2\bar{r}_4-b_1r_3\bar{r}_2\bar{b}_4+g_1b_3\bar{g}_2\bar{b}_4 \nonumber \\
&-g_1b_3\bar{b}_2\bar{g}_4+b_1g_3\bar{b}_2\bar{g}_4-b_1g_3\bar{g}_2\bar{b}_4). \nonumber \\
%\end{align*}
%\begin{align*}
&|C_{4}\rangle = \sqrt{\frac{1}{24}}(2r_1r_3\bar{r}_2\bar{r}_4+2g_1g_3\bar{g}_2\bar{g}_4+2b_1b_3\bar{b}_2\bar{b}_4+r_1g_3\bar{r}_2\bar{g}_4 \nonumber \\
&    +r_1g_3\bar{g}_2\bar{r}_4 +g_1r_3\bar{g}_2\bar{r}_4+g_1r_3\bar{r}_2\bar{g}_4+r_1b_3\bar{r}_2\bar{b}_4 +r_1b_3\bar{b}_2\bar{r}_4\nonumber \\
&+b_1r_3\bar{b}_2\bar{r}_4 +b_1r_3\bar{r}_2\bar{b}_4+g_1b_3\bar{g}_2\bar{b}_4+g_1b_3\bar{b}_2\bar{g}_4+b_1g_3\bar{b}_2\bar{g}_4\nonumber \\
&+b_1g_3\bar{g}_2\bar{b}_4).\nonumber \\
\end{align}

\subsubsection{total wave function}

The total wave functions are obtained by the direct product of wave functions of orbital, spin, color and flavor wave functions. Because we are interested in the states with quantum number $J^P=1^{-}$, there must be orbital angular momentum excitation. The experiment suggests that the excited orbital angular quantum number should exist in one sub-cluster. So we follow the suggestion, set $l_1=1, l_2=0$ or $l_1=0, l_2=1$.  It is worthwhile to mention that it is not easy to calculate matrix elements which relate to excited orbital angular momenta of the systems. In this case, we use infinitesimally shifted Gaussian
method to calculate the matrix elements.

Finally, the total wave function of the \tetraquark ~system is written as:
\begin{equation}
\Psi_{JM_J}^{i,j,k}={\cal A} \left[ \left[  \psi_{L}\chi^{\sigma_i}_{S}\right]_{JM_J} \chi^{fi}_j \chi^{ci}_k \right],
\end{equation}
where the $\cal{A}$ is the antisymmetry operator of the system which guarantees the antisymmetry of the total wave functions when identical particles exchange. Because there are no identical particles in $b\bar{b}q\bar{q}$ \tetraquark~system, the  $\cal{A}$~$=1$. At last, we solve the following Schr\"{o}dinger equation to obtain eigen-energies of the system, with the help of the Rayleigh-Ritz variational principle.
\begin{equation}
H\Psi_{JM_J}^{i,j,k}=E\Psi_{JM_J}^{i,j,k},
\end{equation}
where $\Psi_{JM_J}^{i,j,k}$ is the wave function of the four-quark states, which is the linear combinations of the above channel wave functions.

%\subsubsection{total wave function}
%The total wave functions are obtained by the direct product of wave functions of orbital, spin, color and flavor wave functions.
%Because we are interested in the states with quantum number $J^P=1^{-}$, there must be orbital angular momentum excitation.
%The experiment suggests that the excited angular quantum number should exist in the one sub-cluster. So we follow the suggestion,
%set $l_1=1, l_2=0$ or $l_1=0, l_2=1$. All the possible channels with the physical contents are listed in the table~\ref{channel}.
%The subscript ``8" denotes color octet subcluster, the superscript of diquark/antidiquark is the spin of the subcluster, and the
%subscript is the color representation of subcluster, $3$, $\bar{3}$, $6$ and $\bar{6}$ denote color triplet, anti-triplet,
%sextet and anti-sextet.
\section{stabilization method}
The real-scaling method was first put forward by Taylor \cite{Taylor:1970} to estimate  the energies of long-lived metastable states of electron-atom, electron-molecule, and atom-diatom complexes. Then Jack Simons \cite{simons:1981} adopted this method to study  resonance state. The real-scaling method was firstly applied to quark model by Emiko Hiyama  \emph{et.al.} \cite{Hiyama:2018ukv}  to search for $P_c$ state in the $qqqc\bar{c}$ system. Different from the other computing resonances method based on stabilized eigenvector, the real scaling method can estimate the decay width directly from the stabilization graph in a manner.

\begin{figure}[htp]

  \setlength {\abovecaptionskip} {-0.4cm}
  \centering
  \resizebox{0.50\textwidth}{!}{\includegraphics[width=3cm,height=2.2cm]{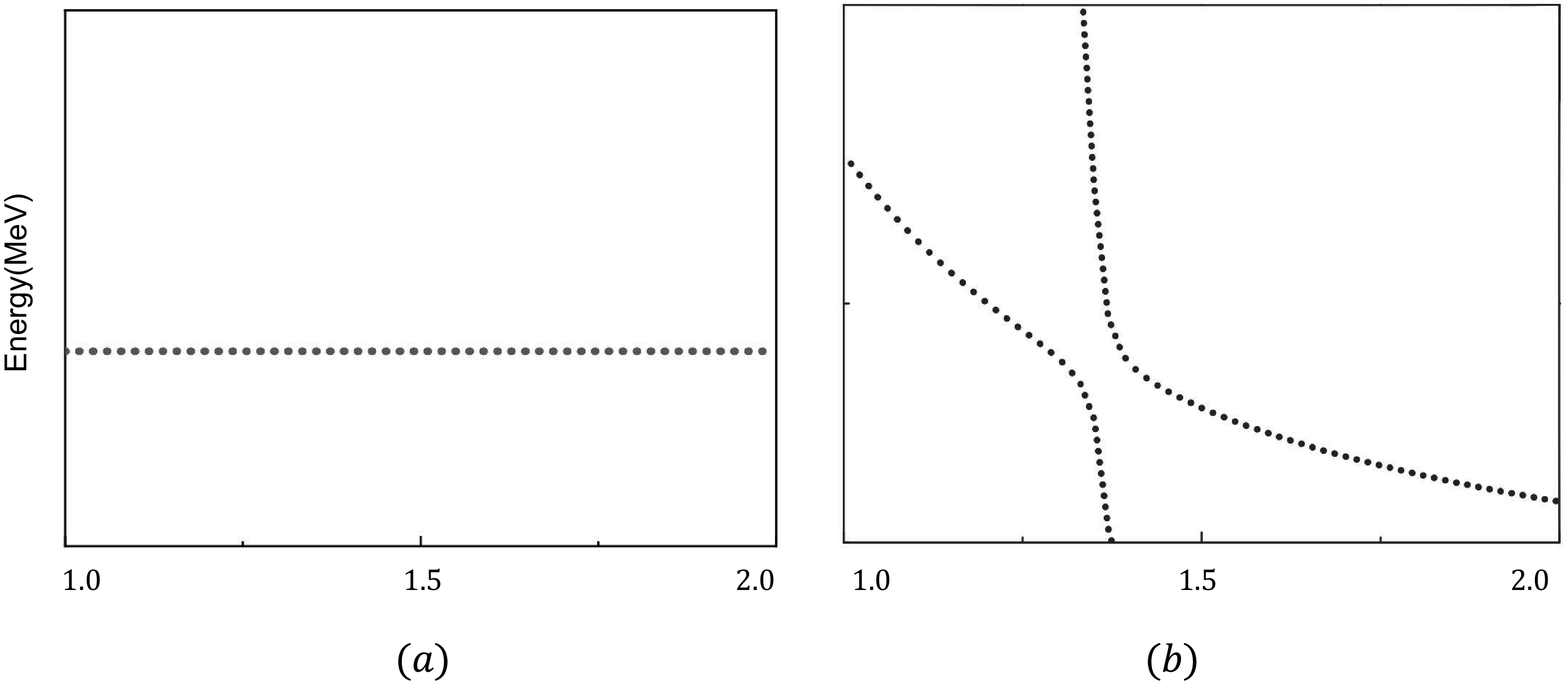}}
  \caption{Two forms of resonant states (a) the resonance has weak coupling (or no coupling) with the scattering states; (b) the resonances has strong coupling with the scattering states;}
\end{figure}
In this approach, a factor $\alpha$ is used to scale the finite volume. The ``false resonant states" reproduced by superabundant  colorful  subclusters (molecular hidden-color state or \da~ state) will fall down to the corresponding threshold, while the genuine  resonances will  survive after coupling to the scattering states and  keep stable with increasing $\alpha$. In this way, the genuine  resonances will have two forms. (a) If the energy of a scattering state is far away from that of the resonance which means there is a weak coupling (or no coupling) between the resonances and the scattering states, the shape of the resonance is a stable straight line; (b) If the energy of a scattering state is close to the resonance, there is a strong coupling between the resonance and the scattering state which would act as an avoid crossing structure between two declining lines.  By this means,  we can estimate the decay width from the slopes of resonance and scattering states from Eq.~(\ref{formula_RSM}), where $S_r$ donates the slope of resonance, $S_s$ donates the slope of scattering state and $\alpha_c$ donates energy level difference between the resonance and the scattering state. In  addition, if $\alpha$ is increasing continually, the avoid crossing structure will repeat again.

\begin{equation}\label{formula_RSM}
\Gamma =4 |V(\alpha_c)|\frac{\sqrt{|s_r||s_s|}}{|s_r-s_s|}
\end{equation}

\section{Results and discussions}
In this section, we present the numerical results of our calculation. Since the newly observed states $\Upsilon(10753)$ can be two-quark or four-quark states, it's necessary to do a comprehensive calculation involving both two-quark and four-quark structure. The calculation results of all mesons from light to heavy used in this paper  are listed in Table \ref{mesons}. For four-quark calculation, the main goal is to search for resonance states which can be candidate of $\Upsilon(10753)$. As the first step, we do a dynamic calculation based on GEM to check whether or not there is any bound state. In our calculation, we expand the wavefunction of two sub-cluster and the wavefunction of relative motion in  Schr\"{o}dinger equation by Gaussian bases. Actually, it's not easy for us to generate Hamilton matrix with spin-orbit interaction. Concerning the spin-orbit interaction has a little effect on  $\chi_{bJ,J=0,1,2}$ meson, as a preliminary calculation, we only take center force without spin-orbit interaction into our consideration at the present four-quark calculation work. So the states can be classified according to the total spin $S$ of the four-quark system, and we mark $S=0$  four-quark system  with $^1P_1$ ($^{2S+1}L_J$), $S=1$  four-quark system  with $^3P_1$ and $S=2$  four-quark system  with $^5P_1$. All the spin of four-quark system $S=0,1,2$ can couple with $L=1$ to give total angular momentum $J=1$. Then we can obtain the energy of the system by solving this generalized eigen-equation. In the present calculation, the meson-meson separation is taken to be less than 6 fm while the quark-antiquark separation in meson is taken to be less than 2 fm.
\subsection{two quark calculation }

%\begin{table}[ht]
%\centering
%\caption{The effect of S-D mixing on the spectrum of $\Upsilon(nS)$ and $\Upsilon(nD)$. We use c.c. denote the result of  coupling of $\Upsilon(nS)    $ and $\Upsilon(nD)$.(unit: MeV). The number n ranges from 4 to 5.\label{SDmix}}
%\begin{tabular}{cccccccccccccccccccccc}
%\hline \hline
%%   c.c.            &  9498 &        &  9967 &       & 10087 &       &  10278 &       &  10361 &       \\
%%$\Upsilon(nS)    $ &  9499 & 100 \% &  9967 & 100\% &  9967 &   0\% &  10278 & 100\% &  10278 &   0\% \\
%%$\Upsilon((n-1)D)$ &   -   &   0 \% & 10087 &   0\% & 10087 & 100\% &  10361 &   0\% &  10361 & 100\% \\
%%\hline
%% 10564 &       &  10631 &       &  10863 &       \\%&  10913 &     \\
%%$\Upsilon(4S)$ 10564 &   99\% &$\Upsilon(4S)$  10564 & 0.1\% &$\Upsilon(5S)$  10863 &   99\% \\%&  10863 &   0\%\\
%%$\Upsilon(3D)$ 10631 &  0.1\% &$\Upsilon(3D)$  10631 &  99\% &$\Upsilon(4D)$  10913 &  0.1\% \\%&  10913 & 100\%  \\
%state   & Mass &   \\
%\hline\hline
%\end{tabular}
%\end{table}

\begin{table}[h]
\caption{\label{SDmix} Mass, in MeV, and probabilities of the different  components, in \%, of the mixing $\Upsilon$ state.}
\begin{ruledtabular}
\begin{tabular}{ccccc}
Mixing State      & Mass  & ${\cal P}[\Upsilon(4S)]$ & ${\cal P}[\Upsilon(3D)]$ \\
$\Upsilon(10564)$ & 10564 & $99.9\%$ &  $0.1\%$ \\ \hline
   Mixing State               &  Mass     & ${\cal P}[\Upsilon(4S)]$ & ${\cal P}[\Upsilon(3D)]$ \\
$\Upsilon(10564)$ & 10631 & $0.1\%$ &  $99.9\%$ \\ \hline
   Mixing State               &    Mass   & ${\cal P}[\Upsilon(5S)]$ & ${\cal P}[\Upsilon(4D)]$ \\
$\Upsilon(10863)$ & 10863 & $99.9\%$ &  $0.1\%$ \\
\end{tabular}
\end{ruledtabular}
\end{table}

%\begin{table}[h]
%\caption{\label{SDmix} Mass, in MeV, and probabilities of the different  components, in \%, of the mixing $\Upsilon$ state.}
%\begin{ruledtabular}
%\begin{tabular}{ccccc}
%Mixing State      & Mass  & ${\cal P}[\Upsilon(4S)]$ & ${\cal P}[\Upsilon(3D)]$ \\
%$\Upsilon(10564)$ & 10564 & $99.9\%$ &  $0.1\%$ \\
%                  &       & ${\cal P}[\Upsilon(4S)]$ & ${\cal P}[\Upsilon(3D)]$ \\
%$\Upsilon(10564)$ & 10631 & $0.1\%$ &  $99.9\%$ \\
%                  &       & ${\cal P}[\Upsilon(5S)]$ & ${\cal P}[\Upsilon(4D)]$ \\
%$\Upsilon(10863)$ & 10863 & $99.9\%$ &  $0.1\%$ \\
%\end{tabular}
%\end{ruledtabular}
%\end{table}

The mass of $\Upsilon(10753)$ observed by Belle Collaboration is $10753\pm6$ MeV. Follow the suggestion from Refs.\cite{Li:2019qsg,Chen:2019uzm} that $\Upsilon(10753)$ may be the mixing of $\Upsilon(4S)$ and $\Upsilon(3D)$ in the quark model, we list our calculated  mass in Table \ref{mesons} and composition information in Table \ref{SDmix}.  We see that the mass of $\Upsilon(4S)$ is 10.56 GeV while the mass of $\Upsilon(3D)$ is 10.63 GeV.  The energy of $\Upsilon(5S)$ is up to 10.8 GeV. In terms of mass, the $\Upsilon(4S)$ is more likely to be the candidate of $\Upsilon(10580)$. Although the S-D mixing effecting may partially ameliorating this difference  between the mass of  $\Upsilon(3D)$ and $\Upsilon(4S)$ in theory, the $\Upsilon(3D)$ of the mixture is unlikely to be the candidate of $\Upsilon(10753)$ due to the mass of  $\Upsilon(3D)$ of the mixture in our work  about 110 MeV less than the mass of $\Upsilon(10753)$. Similarly, the difference in mass prevents $\Upsilon(4S)$ from being a candidate of $\Upsilon(10753)$. On the other hand, the mass of b quark is so  bigger  that the coupling matrix element between $\Upsilon(4S)$ and $\Upsilon(3D)$ would be small, which may lead to weak S-D mixing effecting only about 0.1\%. So, $\Upsilon(10753)$ can not be two quark structure in our work while  the $\Upsilon(10580)$ and $\Upsilon(10860)$ can be traditional meson $\Upsilon(4S)$ and $\Upsilon(5S)$. According to the suggestion from Belle that the $\Upsilon(10753)$ and $\Upsilon(10860)$ may have different substructures, we confirm  the  $\Upsilon(10860)$ is two-quark substructure while the $\Upsilon(10753)$ may have non-$q\bar{q}$  substructure.

\subsection{four quark calculation }
\begin{figure}[!h]
\vspace{-1.2cm}
\setlength {\abovecaptionskip} {-1.0cm}
\setlength {\belowcaptionskip} {-0.2cm}
  \centering
  \includegraphics[width=9cm,height=10cm]{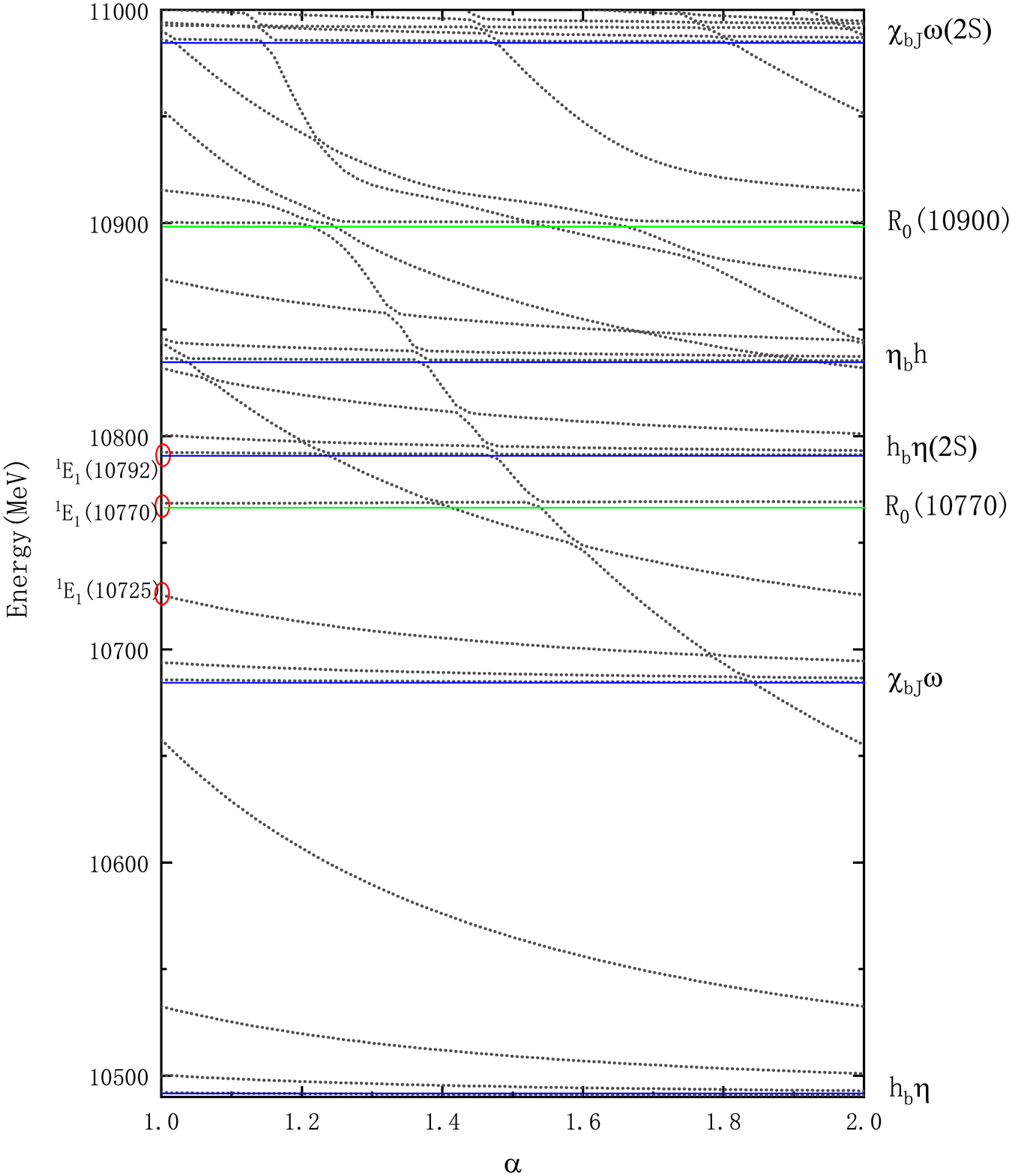}
  \caption{Energy spectrum of $^1P_1$ system.}
   \label{1P1}
\end{figure}

\begin{figure}[!h]
 \vspace{-0.8cm}
  \setlength {\abovecaptionskip} {-0.5cm}
  \setlength {\belowcaptionskip} {-0.2cm}
  \centering
  \includegraphics[width=9cm,height=7cm]{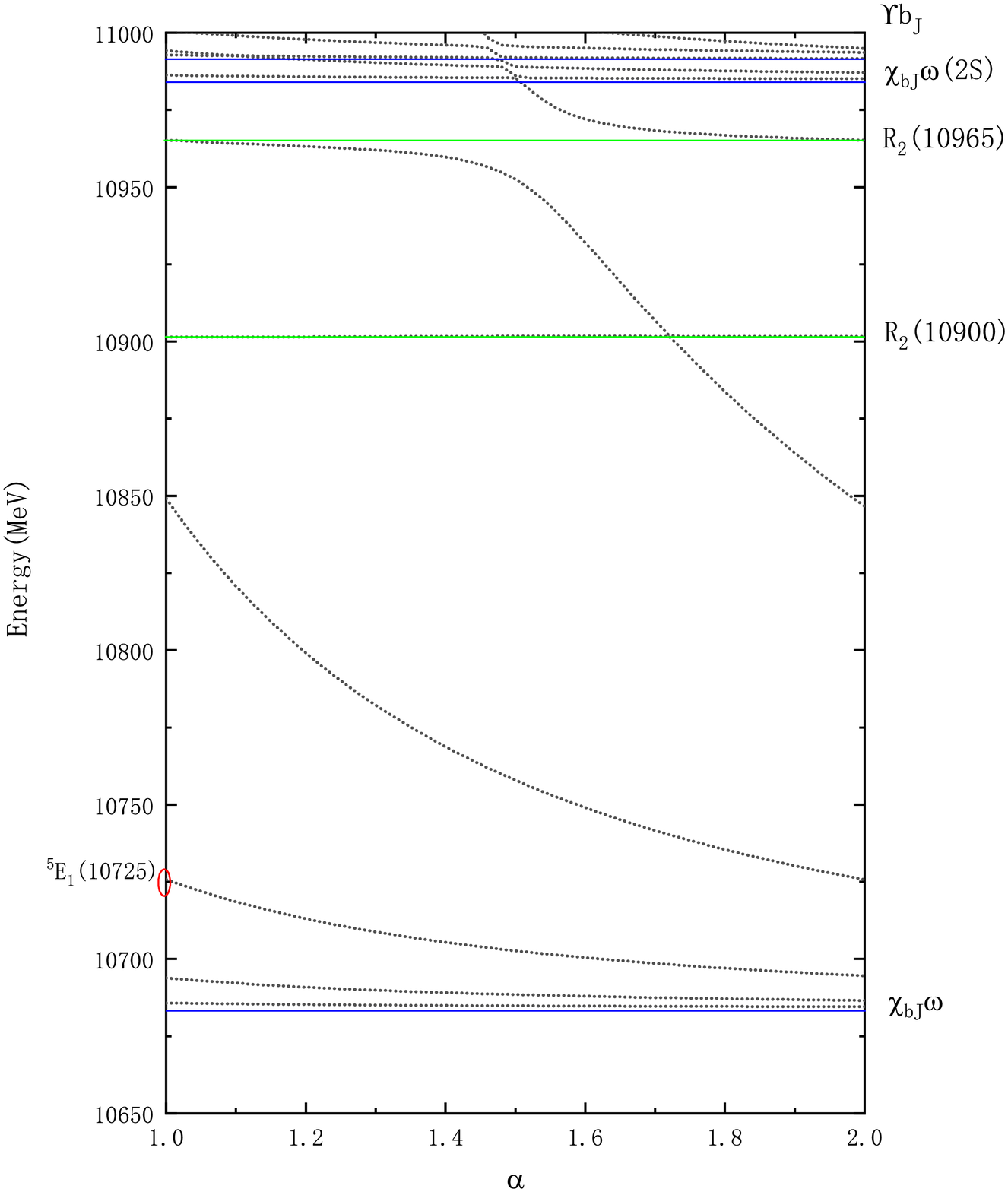}
  \caption{Energy spectrum of $^5P_1$ system.}
 \label{5P1}
\end{figure}

In Tables \ref{S0}, \ref{S1}, \ref{S2}, the significant calculation results are shown. In each table, columns 1 to 4 represent the indices of orbit, flavor, spin and color wave functions in each channel. Column 5 is the corresponding physical channel of tetraquark systems. In column 6, the eigen-energy of each channel is listed and the theoretical threshold (the sum of the theoretical masses of corresponding mesons) is given in column 7. The last column  gives the  experimental thresholds (the sum of the experimental masses of mesons). In the present calculation, we investigate all possible  negative parity $b\bar{q}q\bar{b}$  tetraquarks by taking into account color-singlet di-meson, color-octet di-meson and diquark-antidiquark configurations.  Of these three configurations, color-octet di-meson and diquark-antidiquark configurations can be classified as colorful-subcluster configuration, of which every eigen-energy is resonant state.  At the bottom of every structure  configuration is the result of their channel coupling, where the lowest-lying energy is listed. When a complete coupled-channels calculation is performed, last row of Table \ref{S0}, \ref{S1}, \ref{S2} indicates the lowest-lying mass of tetraquark systems. When the RSM is used in the complete coupled-channels calculation shown in Figs. \ref{5P1} to \ref{1P1}, the avoid-crossing points can be obtained in the distribution of eigen-energies  of very channel, which indicate the possible resonances.

\begin{table*}[htb]
\caption{The results for system with $^{2S+1}L_J$=$^1P_1$. The ``$[meson]_8$'' donates molecular color-octet state. ``$[sub$-$diquark]_{color}^{spin}$'' denotes diquark state, and  ``$[sub$-$antidiquark]_{anti-color}^{spin}$'' denotes antidiquark state. The calculated eigen-energies closest to $\Upsilon(10753)$ are listed at the bottom of table, which denoted by $^{2S+1}E_{J}(eigen$-$energy)$.(unit: MeV)\label{S0}}
\begin{tabular}{cccccccccc}
\hline \hline
~~~~~$|R_i\rangle$~~~~~~&~~~~~$|F_j\rangle$~~~~~~&~~~~~$|S_k\rangle$~~~~~&~~~~~$|C_n\rangle$~~~~~&~~~~Channel~~~~~&~~~E~~~~&~~~$E^{Theo}_{th}$~~~~&
~~~$E_B$~~~&~~~$E^{Exp}_{th}$~~~\\

\hline
 $i=1$ &$j=1$ & $k=1$ & $n=1$ & $B_1^{'}B$                            &  11105&  11104&0  & 11005 \\%&      \\
 $i=1$ &$j=1$ & $k=2$ & $n=1$ & $B_JB^*$                              &  11169&  11168&0  & 11056\\%&      \\
 $i=1$ &$j=2$ & $k=1$ & $n=1$ & $h_b\eta$                             &  10492&  10489&0  & 10447\\%&      \\
 $i=1$ &$j=2$ & $k=2$ & $n=1$ & $\chi_{bJ}\omega$                     &  10685&  10684&0  & 10675\\%&      \\
 $i=2$ &$j=2$ & $k=2$ & $n=1$ & $\Upsilon b_J$                        &  10836&  10834&0  & 10741\\%&      \\
 $i=2$ &$j=2$ & $k=1$ & $n=1$ & $\eta_b h$                            &  10992&  10990&0  & 10565\\%&      \\
 \multicolumn{5}{c}{coupled-meson-channels:}                          &  10492\\
  \\
 $i=1$ &$j=1$ & $k=1$ & $n=2$ & $[B_1^{'}]_8[B]_8$                    &  11148&  11104&0  & 11005 \\%&      \\
 $i=1$ &$j=1$ & $k=2$ & $n=2$ & $[B_J]_8[B^*]_8$                      &  11099&  11168&0  & 11056\\%&      \\
 $i=1$ &$j=2$ & $k=1$ & $n=2$ & $[h_b]_8[\eta]_8$                     &  11103&  10489&0  & 10447\\%&      \\
 $i=1$ &$j=2$ & $k=2$ & $n=2$ & $[\chi_{bJ}]_8[\omega]_8$             &  10982&  10684&0  & 10675\\%&      \\
 $i=2$ &$j=2$ & $k=2$ & $n=2$ & $[\Upsilon]_8[b_J]_8$                 &  11342&  10834&0  & 10741\\%&      \\
 $i=2$ &$j=2$ & $k=1$ & $n=2$ & $[\eta_b]_8[h]_8$                     &  11299&  10990&0  & 10565\\%&      \\
 $i=3$ &$j=3$ & $k=1$ & $n=3$ & $[bq]_3^0[\bar{b}\bar{q}]_{\bar{3}}^0$&  11312&  $-$  &$-$& $-$\\%&      \\
 $i=3$ &$j=3$ & $k=1$ & $n=4$ & $[bq]_6^0[\bar{b}\bar{q}]_{\bar{6}}^0$&  11240&  $-$  &$-$& $-$\\%&      \\
 $i=3$ &$j=3$ & $k=2$ & $n=3$ & $[bq]_3^1[\bar{b}\bar{q}]_{\bar{3}}^1$&  11339&  $-$  &$-$& $-$\\%&      \\
 $i=3$ &$j=3$ & $k=2$ & $n=4$ & $[bq]_6^1[\bar{b}\bar{q}]_{\bar{6}}^1$&  11202&  $-$  &$-$& $-$\\%&      \\
\multicolumn{5}{c}{coupled-colorful-subcluster-channels:}              &  10813\\
  \\
  \multicolumn{5}{c}{Complete coupled-all-channels:}                   & $^1E_1(10725)$& $^1E_1(10770)$&   $^1E_1(10792)$ \\
\hline \hline
\end{tabular}
\end{table*}

\begin{figure}[h]
\vspace{-1.3cm}
\setlength {\abovecaptionskip} {-1.4cm}
\setlength {\belowcaptionskip} {-0.4cm}
  \centering
  \includegraphics[width=9cm,height=13cm]{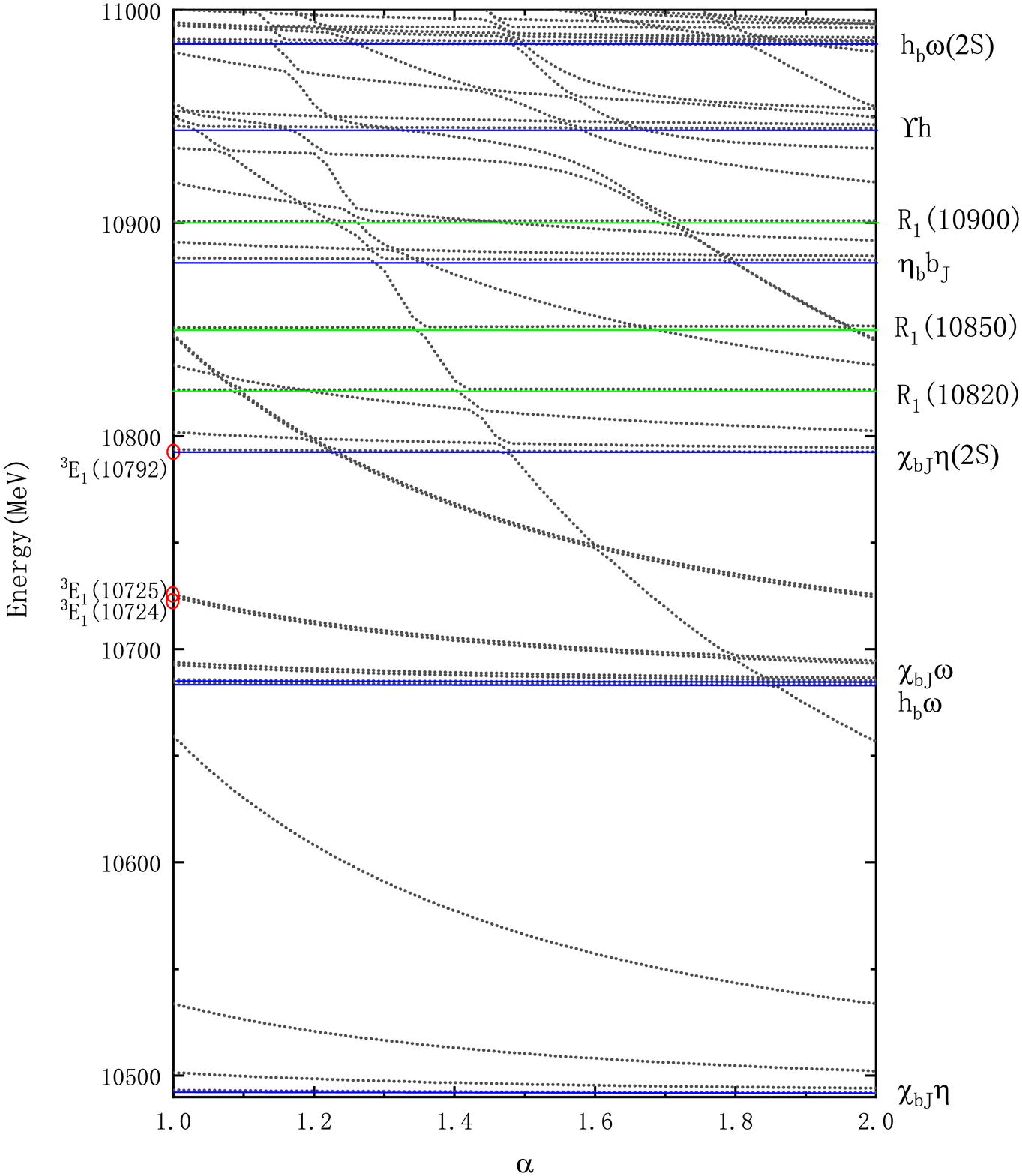}
  \caption{Energy spectrum of $^3P_1$ system.}
   \label{3P1}.
\end{figure}

\begin{table*}[th]
\caption{The main component and root-mean-square distances of candidates of $\Upsilon(10753)$ in all systems.\label{rms}}
\begin{tabular}{cccccccccccccccccccccccccccccccccccccccccccccc} \hline \hline
State&~~  & \multicolumn{3}{c}{Main component}  & &$r_{b\bar{q}}$&~ &$r_{bq}$&~  &$r_{b\bar{b}}$&~  &$r_{\bar{q}q}$&    &$r_{\bar{q}\bar{b}}$&~& $r_{q\bar{b}}$&\\ \hline
$^1E_1(10725)$&  & \multicolumn{3}{c}{$\chi_{bJ}\omega$($97\%$) + the others.}                        & &5.66& &5.66&  &0.39& &0.51&  &5.66& &5.66&\\
$^1E_1(10770)$&  & \multicolumn{3}{c}{diquark-antidiqaurk($45\%$)+color-octet($37\%$) + the others.}  & &0.75& &0.78&  &0.34& &0.60&  &0.78& &0.76&\\
$^1E_1(10792)$&  & \multicolumn{3}{c}{$h_b\eta$($98\%$) + the others.}                                & &5.66& &5.66&  &0.38& &0.46&  &5.66& &5.66&\\
$^3E_1(10725)$&  & \multicolumn{3}{c}{$h_b\omega$($97\%$) + the others.}                              & &5.66& &5.66&  &0.38& &0.51&  &5.66& &5.66&\\
$^3E_1(10768)$&  & \multicolumn{3}{c}{$\chi_{bJ}\omega $($98\%$ + the others.}                        & &5.66& &5.66&  &0.39& &0.51&  &5.66& &5.66&\\
$^3E_1(10792)$&  & \multicolumn{3}{c}{$\chi_{bJ}\eta $($98\%$) + the others.}                         & &5.66& &5.66&  &0.39& &0.46&  &5.66& &5.66&\\
$^5E_1(10725)$&  & \multicolumn{3}{c}{$\chi_{bJ}\omega $($98\%$) + the others.}                       & &5.66& &5.66&  &0.39& &0.51&  &5.66& &5.66&\\
 \hline\hline
\end{tabular}
\end{table*}

\begin{table*}[htb]
\caption{The results for system with $^{2S+1}L_J$=$^5P_J$. The ``$[meson]_8$'' denotes molecular color-octet state. ``$[sub$-$diquark]_{color}^{spin}$'' denotes diquark state, and ``$[sub$-$antidiquark]_{anti-color}^{spin}$'' denotes antidiquark state.  The calculated eigen-energies closest to $\Upsilon(10753)$ are listed at the bottom of table, which denoted by $^{2S+1}E_{J}(eigen$-$energy)$.(unit: MeV)\label{S2}}
\begin{tabular}{ccccccccccc}
\hline \hline
~~~~~$|R_i\rangle$~~~~~~&~~~~~$|F_j\rangle$~~~~~~&~~~~~$|S_k\rangle$~~~~~&~~~~~$|C_n\rangle$~~~~~&~~~~Channel~~~~~&~~~E~~~~&~~~$E^{Theo}_{th}$~~~~&
~~~$E_B$~~~&~~~$E^{Exp}_{th}$~~~\\

\hline
 $i=1$ &$j=1$ & $k=6$ & $n=1$ & $B_JB^*$                              & 11169 & 11168 &0 &  11056    &      \\
 $i=1$ &$j=2$ & $k=6$ & $n=1$ & $\chi_{bJ}\omega$                     & 10685 & 10684 &0 &  10675    &      \\
 $i=2$ &$j=2$ & $k=6$ & $n=1$ & $\Upsilon b_J$                        & 10992 & 10990 &0 &  10741    &      \\
 \multicolumn{5}{c}{Complete coupled-meson-channels}                 &  10685 \\
  \\
 $i=1$ &$j=1$ & $k=6$ & $n=2$ & $[B_J]_8[B^*]_8$                      & 11166 & 11168 &$-$  & $-$     &      \\
 $i=1$ &$j=2$ & $k=6$ & $n=2$ & $[\chi_{bJ}]_8[\omega]_8$            & 11029 & 10684 &$-$  & $-$     &      \\
 $i=2$ &$j=2$ & $k=6$ & $n=2$ & $[\Upsilon]_8[b_J]_8$                 & 11328 & 10990 &$-$  & $-$     &      \\
 $i=3$ &$j=3$ & $k=6$ & $n=3$ & $[bq]_3^1[\bar{b}\bar{q}]_{\bar{3}}^1$& 11332 & $-$   & $-$ & $-$     &      \\
 $i=3$ &$j=3$ & $k=6$ & $n=4$ & $[bq]_6^1[\bar{b}\bar{q}]_{\bar{6}}^1$& 11247 & $-$   & $-$ & $-$     &      \\
\multicolumn{5}{c}{Complete coupled-colorful-subclusters-channels}   & 10935 \\
  \\
  \multicolumn{5}{c}{Complete coupled-all-channels}                 &  $^5E_1(10725)$ \\

\hline \hline
\end{tabular}
\end{table*}

$\bf{The~^{2S+1}L_{J}=^{1}P_1~channel}:$ There are six color-singlet channels, $B_1^{'}B$, $B_JB^{*}$, $h_b\eta$, $\chi_{bJ}\omega$, $\Upsilon b_J$ and $\eta_bh$ included in the  di-meson configuration  while the ``colorful-subcluster" configuration includes six color-octet channels  along wih  six diquark-antidiquark channels  individually shown in Table \ref{S0}. The result indicates every color-singlet state is scattering state and the energy of each channel locates above the corresponding threshold channel $h_b\eta(10447)$. We perform a channel-coupling for all color-singlet states, and find no bound state in $^1P_1$ system. Then in the fully channel-coupling calculation of the ``colorful-subcluster" configuration,  the calculated masses of color-octet channels and diquark-antidiquark channels are located in an interval 10.9-11.2 GeV. We notice that although the color-singlet mass of the $h_b\eta$ is smaller than that of the $\chi_{bJ}\omega$, the color-octet mass of the $[\chi_{bJ}]_8[\omega]_8$ is smaller than that of the $[h_b]_8[\eta]_8$. The possible reason for this phenomenon is that there is more attraction between two vector mesons than between two scalars where the one-gluon exchange and kinetic energy may play an important role Seen in Ref.\cite{Wu:2021ahn}.  When  partially-coupled calculation is performed for the ``colorful-subcluster" configuration, the calculated mass of $[\chi_{bJ}]_8[\omega]_8$ is pushed down to 10813 MeV, which is very close to the mass of $\Upsilon(10753)$. However, the state with the lowest energy of the ``colorful-subcluster" configuration spectrum is not necessarily stable. It should undergo the coupling with decay channels which could put up the resonances to be scattering state (Seen in Ref.\cite{Hu:2022pae}, the resonant $0\frac{1}{2}^{-}$ $\Xi_c\bar{K}^{*}$  decays strongly to the open channels) and helps the unbound molecular state form a bound state (Seen in Table 6 in Ref.\cite{Chen:2021tnn}, the $DD^{*}$ becomes a bound state with the help of $[D]_8[D^{*}]_8$). Thus, the complete coupled-all-channels is performed, and several energies near the $\Upsilon(10750)$ range from 10.7 to 10.8 GeV are obtained, $^1E_1(10725)$, $^1E_1(10768)$ and $^1E_1(10792)$ (Herein we use $^{2S+1}E_{J}(eigen$-$energy)$ denotes the eigen-value of complete coupled-all-channels). We have investigated the composition of these states in Table \ref{rms}. For $^1E_1(10725)$ and $^1E_1(10792)$, the dominant structures are scattering states $\chi_{bJ}\omega$ and $h_b\eta(2S)$  occupying  the vast majority of the components of the total wave function, which indicates the fully coupling energies $^1E_1(10725)$ and $^1E_1(10792)$  may not be stable resonances.  Table \ref{rms} also show that the distance between $b$ quark and $\bar{b}$ quark is 0.3 fm, and the other quarks are all around 0.7 fm. The distances indicates the $^1E_1(10770)$ may be a compact tetraquark configuration which is correspondence  with our calculated result that the $^1E_1(10770)$ has 45\% diquark-antidiquark and 37\% color-octet composition. When the calculation is performed in the framework of Real-scaling method, we can find that the $^1E_1(10725)$ and $^1E_1(10792)$ rapidly  fall down to the corresponding threshold seen the red circle on the left panel of Fig.\ref{1P1} while the $^1E_1(10770)$ acts as a straight green line, which indicates that the $^1E_1(10770)$ may be  a stable resonance ($R_0(10770)$, Herein we use $^{2S+1}R_{J}(eigen$-$energy)$ denotes the stable resonance with a definite spin).  Besides, another green straight line is also obtained with a mass of 10.9 GeV, which can be named as $^1R_1(10900)$.

\begin{table}[ht]
\centering
\caption{The width of candidates of $\Upsilon(10753)$. Herein we use ``$^{2S+1}R_{J}(eigen$-$energy)$" denotes the stable Resonance with a definite spin.(unit: MeV).\label{width}}
\begin{tabular}{cccccccc}
\hline \hline
state ~~& $^1R_1(10770)$~~  & $^1R_1(10900)$ ~~& $^3R_1(10820)$  ~~&  $^3R_1(10850)$   \\
width ~~&  16           &    0.1           &   1.7          &  2.8        \\
state~~&  $^3R_1(10900)$  &  $^5R_1(10900)$ &  $^5R_1(10965)$  \\
width ~~&  0.6           &  0.7           & 160   \\

\hline\hline
\end{tabular}
\end{table}
\begin{table*}[htb]
\caption{The results for system with $^{2S+1}L_J$=$^3P_J$. The ``$[meson]_8$'' denotes molecular color-octet state. ``$[sub$-$diquark]_{color}^{spin}$'' denotes diquark state while ``$[sub$-$antidiquark]_{anti-color}^{spin}$'' denotes antidiquark state. The calculated eigen-energies closest to $\Upsilon(10753)$ are listed at the bottom of table, which denoted by $^3E_{J}(eigen$-$energy)$.(unit: MeV)\label{S1}}
\begin{tabular}{ccccccccccc}
\hline \hline
~~~~~$|R_i\rangle$~~~~~~&~~~~~$|F_j\rangle$~~~~~~&~~~~~$|S_k\rangle$~~~~~&~~~~~$|C_n\rangle$~~~~~&~~~~Channel~~~~~&~~~E~~~~&~~~$E^{Theo}_{th}$~~~~&
~~~$E_B$~~~&~~~$E^{Exp}_{th}$~~~\\

\hline
 $i=1$ &$j=1$ & $k=1$ & $n=1$ & $B_1^{'}B^*$                          &  11162&  11161 &0 & 11051\\
 $i=1$ &$j=1$ & $k=2$ & $n=1$ & $B_JB$                                &  11112&  11111 &0 & 11011\\
 $i=1$ &$j=1$ & $k=3$ & $n=1$ & $B_JB^*$                              &  11169&  11168 &0 & 11057\\
 $i=1$ &$j=2$ & $k=1$ & $n=1$ & $h_b\omega$                           &  10684&  10682 &0 & 10681\\
 $i=1$ &$j=2$ & $k=2$ & $n=1$ & $\chi_{bJ}\eta$                       &  10493&  10491 &0 & 10441\\
 $i=1$ &$j=2$ & $k=3$ & $n=1$ & $\chi_{bJ}\omega$                     &  10685&  10684 &0 & 10674\\
 $i=2$ &$j=2$ & $k=1$ & $n=1$ & $\eta_b b_J$                          &  10883&  10881 &0 & 10679\\
 $i=2$ &$j=2$ & $k=2$ & $n=1$ & $\Upsilon h$                          &  10945&  10943 &0 & 10626\\
 $i=2$ &$j=2$ & $k=3$ & $n=1$ & $\Upsilon b_J$                        &  10992&  10990 &0 & 10741\\
 \multicolumn{5}{c}{Complete coupled-meson-channels}                  &  10493\\
  \\
 $i=1$ &$j=1$ & $k=1$ & $n=2$ & $[B_1^{'}]_8[B^*]_8$                  &  11145&  11161 &0 & 11051\\
 $i=1$ &$j=1$ & $k=2$ & $n=2$ & $[B_J]_8[B]_8$                        &  11148&  11111 &0 & 11011\\
 $i=1$ &$j=1$ & $k=3$ & $n=2$ & $[B_J]_8[B^*]_8$                      &  11122&  11168 &0 & 11057\\
 $i=1$ &$j=2$ & $k=1$ & $n=2$ & $[h_b]_8[\omega]_8$                   &  11014&  10682 &0 & 10681\\
 $i=1$ &$j=2$ & $k=2$ & $n=2$ & $[\chi_{bJ}]_8[\eta]_8$               &  11103&  10491 &0 & 10441\\
 $i=1$ &$j=2$ & $k=3$ & $n=2$ & $[\chi_{bJ}]_8[\omega]_8$             &  10998&  10684 &0 & 10674\\
 $i=2$ &$j=2$ & $k=1$ & $n=2$ & $[\eta_b]_8[b_J]_8$                   &  11319&  10881 &0 & 10679\\
 $i=2$ &$j=2$ & $k=2$ & $n=2$ & $[\Upsilon_b]_8[h]_8$                 &  11341&  10943 &0 & 10626\\
 $i=2$ &$j=2$ & $k=3$ & $n=2$ & $[\Upsilon]_8[b_J]_8$                 &  11309&  10990 &0 & 10741\\
 $i=3$ &$j=3$ & $k=1$ & $n=3$ & $[bq]_3^0[\bar{b}\bar{q}]_{\bar{3}}^1$& 11331 &   $-$    & $-$ &  $-$    \\
 $i=3$ &$j=3$ & $k=1$ & $n=4$ & $[bq]_6^0[\bar{b}\bar{q}]_{\bar{6}}^1$& 11234 &   $-$    & $-$ &  $-$    \\
 $i=3$ &$j=3$ & $k=2$ & $n=3$ & $[bq]_3^1[\bar{b}\bar{q}]_{\bar{3}}^0$& 11315 &   $-$    & $-$ &  $-$    \\
 $i=3$ &$j=3$ & $k=2$ & $n=4$ & $[bq]_6^1[\bar{b}\bar{q}]_{\bar{6}}^0$& 11239 &   $-$    & $-$ &  $-$    \\
 $i=3$ &$j=3$ & $k=3$ & $n=3$ & $[bq]_3^1[\bar{b}\bar{q}]_{\bar{3}}^1$& 11337 &   $-$    & $-$ &  $-$    \\
 $i=3$ &$j=3$ & $k=3$ & $n=4$ & $[bq]_6^1[\bar{b}\bar{q}]_{\bar{6}}^1$& 11217 &   $-$    & $-$ &  $-$    \\
 \multicolumn{5}{c}{Complete coupled-colorful-subclusters-channels}   & 10856 \\
  \\
  \multicolumn{5}{c}{Complete coupled-all-channels}                   &  $^3E_1(10724)$& $^3E_1(10725)$&   $^3E_1(10793)$\\

\hline \hline
\end{tabular}
\end{table*}

$\bf{The~^{2S+1}L_{J}=^{3}P_J~channel}:$There are 24 physical channels contributing to the $^3P_J$ $b\bar{q}q\bar{b}$ tetraquark system. Nine molecular color-singlet states, $B_1^{'}B^*$, $B_JB$, $B_JB^*$, $h_b\omega$, $\chi_{bJ}\eta$, $\chi_{bJ}\omega$, $\eta_b b_J$, $\Upsilon h$ and $\Upsilon b_J$, and  nine molecular hidden-color states, along with six diquark-antidiquark states are individually studied in Table \ref{S1}.  The result  shows the obtained masses in each channel  are always above the lowest di-meson threshold. The coupling result of nine molecular color-singlet states indicate that the lowest energy $10.49$ GeV of the system is still located above the lowest-lying threshold of $^{2S+1}L_{J}=^3P_J$ $b\bar{q}q\bar{b}$ tetraquark states. Then we do the partially-coupled calculations for the colorful-subclusters configuration, and the result shows the mass of  the calculated energies  range from 10.9 GeV to 11.3 GeV as well. Because the mass of b quark is so big that color-magnetic interaction  could not separate the masses of  $h_b$ particle and $\chi_{b1}$ particle in our chiral quark model, the energy of $\chi_{bJ}\omega$ is very close to that of $h_b\omega$ Seen in Table \ref{mesonmass}. As in the case of $^1P_1$ $b\bar{q}q\bar{b}$ tetraquark system, although the color-singlet masses of $\chi_{bJ}\omega$ and $h_b\omega$ are similar, $\chi_{bJ}\omega$'s color-octet mass is smaller due to the greater attraction between the di-vector mesons seen in 15 row and 17 row of Table \ref{S1}. In the complete coupled-all-channels calculation, we have noticed that only three energies $^3E_1(10724)$, $^3E_1(10725)$ and $^3E_1(10793)$ near the $\Upsilon(10753)$ shown in Table \ref{S1}.  One can see that these possible candidates are of scattering nature with more than 97\% open channel. The complete coupled-all-channels calculation has been extended to more calculated space by real-scaling method. We have noticed that the three energies $^3E_1(10724)$, $^3E_1(10725)$ and $^3E_1(10793)$ may have strong coupling effect with $\chi_{bJ\omega}$ and $h_b\omega$ thresholds (Seen the candidate curves marked by the red circle drop rapidly to blue threshold lines), while other three energy levels, $^3R_1(10900)$, $^3R_1(10850)$ and $^3R_1(10820)$ curves keep stable where we have marked them by green lines, which indicates all of them may be stable resonances.

$\bf{The~^{2S+1}L_{J}=^{5}P_J~channel}:$Table \ref{S2} lists three color-singlet states and three hidden-color states and two diquark-antidiquark states of the $^5P_J$ $b\bar{q}q\bar{b}$ system.  Every single color-singlet channel is scattering state and  no bound state is found in color-singlet meson channels  coupling result. Because the physical channel number of $^5P_J$ $b\bar{q}q\bar{b}$ system is smaller than $^1P_1$ $b\bar{q}q\bar{b}$ and $^3P_J$ $b\bar{q}q\bar{b}$, the calculated coupled mass in hidden-color, diquark-antidiquark states are bigger than the other systems. If one considers a fully coupled-channel calculation, the lowest mass of $^5P_J$ $b\bar{q}q\bar{b}$ system is located above the threshold of $\chi_{bJ}\omega$, and only one energy level near $\Upsilon(10753)$ is achieved, $^5E_1(10725)$. Obviously, the energy level $^5E_1(10725)$ is a scattering $\chi_{bJ}\omega$ state, similar to the case of $^3E_1(10725)$. In the fully coupled-channels calculation based on real-scaling method, we can see the line of $^5E_1(10725)$  is falling down to the corresponding threshold, which proves the  $^5E_1(10725)$  is of a scattering state. We have noticed that there are two stable green lines in Fig.\ref{5P1}, which indicates other two stable resonances $^5R_1(10900)$ and $^5R_1(10965)$ existing in the $^5P_J$ $b\bar{q}q\bar{b}$ system.

The calculated width of these resonances are listed in Table \ref{width}. One can see the width of $^1R_1(10770)$ is 16 MeV, which is close to that of $\Upsilon(10753)$ ($\Gamma$=$36_{-12}^{+18}$ MeV) within the allowed error range. Thus it's rational to interpret  $^1R_1(10770)$ as experimental $\Upsilon(10753)$ in terms of their  mass and width. Besides, we also obtain the other several resonances $^1R_1(10900)$, $^3R_1(10820)$, $^3R_1(10850)$, $^3R_1(10900)$, $^5R_1(10900)$ and  $^5R_1(10965)$.  According to Table \ref{width}, we have noticed that although the three resonances  $^1R_1(10900)$, $^3R_1(10900)$ and $^5R_1(10900)$ have different spin, they share similar mass (near 10900 MeV) and width ($<$ 1 MeV). As a consequence, we believe that these resonances may be degenerate states, where we need spin-orbit interaction to separate them. For the same reason, the two resonances  $^3R_1(10820)$ and  $^3R_1(10850)$ with similar widths ($<$ 3 MeV) also can be degenerate states. So considering the effect of all possible spin-orbit coupling  is our further study. Because it's very difficult for us to consider the decay process of $^3R_1(10850)$ and $^5R_1(10850)$ $\rightarrow$ $\Upsilon(nS)\pi^{+}\pi^{-}$, we can not deny that   $^3R_1(10850)$ and $^5R_1(10850)$  are not candidates of experimental $\Upsilon(18650)$ ($\Gamma$=$37\pm{5}$ MeV) in term of their smaller width.

\section{Summary}

In the framework of a chiral constituent quark model, we study systematically $J^P=1^{-}$ $\Upsilon(10753)$ state. The two quark $\Upsilon(nS)$ states and four quarks $J^P=1^{-}$ $b\bar{b}q\bar{q}$ are investigated with the help of Gaussian expansion method.

In the result of two-quark system, we notice that the mass of  fourth radial excited state (4S) of $\Upsilon$ is 10564 MeV which is a good candidate of $\Upsilon(10580)$ observed by experiment while the fifth radial excited state (5S) with an  energy of 10863 MeV  is possible candidate of experimental $\Upsilon(10860)$. We find that several authors suggested the S-D mixing may be helpful for increasing the mass of   $3D$ of $\Upsilon$ state  to be candidate of $\Upsilon(10753)$. However, in our theoretical calculation, the S-D mixing is smaller that the mass of the mixed $3D$ of $\Upsilon$ is no different from the original mass.  There are two reasons that the mass of b quark is very big that the cross matrix element is small, and the mass difference between $\Upsilon(4S)$ and $\Upsilon(3D)$ is relatively large.

For the four-quark system, two different meson-meson structure $b\bar{b}$-$q\bar{q}$ and $b\bar{q}$-$q\bar{b}$, one diquark-antidiquark structure $bq$-$\bar{q}\bar{b}$, with all possible color, flavor, spin configurations are taken into account. Firstly, no bound state is found in every color-singlet physical channel, also in the all of color-singlet physical channel coupling. By means of coupling all of colorful-subclusters states, several resonances are achieved in $^{2S+1}L_{J}=^1P_1,^3P_J,^5P_J$ $b\bar{q}q\bar{b}$ systems. Then, to confirm these resonances are the ones that are observed experimentally, ?We check whether they can survive coupling to scattering states in the framework of the real-scaling approach. In the absence of spin-orbit interaction, we rule out several false states which are of scattering nature and obtain one reliable resonance, $^1R_{1}(10770)$ (We use $^{2S+1}R_{J}(mass)$ to donate resonances in different systems with specific spin). According to the fact that our calculated mass and width of $^1R_{1}(10770)$ very close to that of experimental $\Upsilon(10753)$ within the error range, we consider $^1R_{1}(10770)$ with great colorful sub-clusters composition  is a good candidate of $\Upsilon(10753)$. Besides, we obtain the other six resonances $^1R_1(10900)$, $^3R_1(10820)$, $^3R_1(10850)$, $^3R_1(10900)$, $^5R_1(10900)$ and  $^5R_1(10965)$. After comparing their mass and width, we suggest that three resonances $^1R_1(10900)$, $^3R_1(10900)$ and $^5R_1(10900)$ with the same mass belonging to different spins  degenerate states, which need experiments to confirm them in the future.

 %and we need spin-orbit force to distinguish them in further step. %As for the $^3R_1(10820)$ and $^3R_1(10850)$,

According to  the above discussion, the newly observed state $\Upsilon(10753)$ can be described as a compact tetraquark state $^1R_{1}(10770)$ in our work. As for experimental $\Upsilon(10860)$, we can not tell directly whether it's a two-quark structure or a four-quark structure based on our present work due to the lack of decay width calculation ($\Upsilon(nS)\pi^{+}\pi^{-}$). In addition, several resonances are found with energies located in an interval 10.90-10.96 GeV, which corresponds to the suggestion from Belle that no significant signal is observed for masses between 10.45 and 10.65 GeV. When the spin-orbit and tensor interactions are included, all the states with $^1P1$, $^3P_J$ and $^5P_J$ will be mixed. Clearly, further calculation is expected.

\acknowledgments{This work is supported partly by the National Natural Science Foundation of China under grant no. (12205249), by the Natural Science Foundation of Jiangsu Province (BK20221166), and the Funding for School-Level Research Projects of Yancheng Institute of Technology (No. xjr2022039). }


\begin{thebibliography}{99}

\bibitem{Choi:2003}
  S.K.~Choi et al, (Belle Collaboration),
  % Belle firstly found X3872
  Phys.\ Rev.\ Lett {\bf 91}, 262001 (2003).

%\cite{Belle-II:2022xdi}
\bibitem{Belle-II:2022xdi}
I.~Adachi \textit{et al.} [Belle-II],
%``Observation of $e^+e^-\to\omega\chi_{bJ}(1P)$ and search for $X_b \to \omega\Upsilon(1S)$ at $\sqrt{s}$ near 10.75 GeV,''
[arXiv:2208.13189 [hep-ex]].
%0 citations counted in INSPIRE as of 02 Oct 2022


%\cite{Belle:2011aa}
\bibitem{Belle:2011aa}
A.~Bondar \textit{et al.} [Belle],
%``Observation of two charged bottomonium-like resonances in Y(5S) decays,''
Phys. Rev. Lett. \textbf{108}, 122001 (2012)
doi:10.1103/PhysRevLett.108.122001
[arXiv:1110.2251 [hep-ex]].
%625 citations counted in INSPIRE as of 10 Aug 2022

%\cite{Belle:2015upu}
\bibitem{Belle:2015upu}
A.~Garmash \textit{et al.} [Belle],
%``Observation of Zb(10610) and Zb(10650) Decaying to B Mesons,''
Phys. Rev. Lett. \textbf{116}, no.21, 212001 (2016)
doi:10.1103/PhysRevLett.116.212001
[arXiv:1512.07419 [hep-ex]].
%92 citations counted in INSPIRE as of 10 Aug 2022


%\cite{Chen:2019uzm}
\bibitem{Chen:2019uzm}
B.~Chen, A.~Zhang and J.~He,
%``Bottomonium spectrum in the relativistic flux tube model,''
Phys. Rev. D \textbf{101}, no.1, 014020 (2020)
doi:10.1103/PhysRevD.101.014020
[arXiv:1910.06065 [hep-ph]].
%13 citations counted in INSPIRE as of 02 Oct 2022



%\cite{Li:2019qsg}
\bibitem{Li:2019qsg}
Q.~Li, M.~S.~Liu, Q.~F.~L\"u, L.~C.~Gui and X.~H.~Zhong,
%``Canonical interpretation of $Y(10750)$ and $\Upsilon(10860)$ in the $\Upsilon$ family,''
Eur. Phys. J. C \textbf{80}, no.1, 59 (2020)
doi:10.1140/epjc/s10052-020-7626-2
[arXiv:1905.10344 [hep-ph]].
%14 citations counted in INSPIRE as of 02 Oct 2022

%\cite{Wang:2019veq}
\bibitem{Wang:2019veq}
Z.~G.~Wang,
%``Vector hidden-bottom tetraquark candidate: $Y(10750)$,''
Chin. Phys. C \textbf{43}, no.12, 123102 (2019)
doi:10.1088/1674-1137/43/12/123102
[arXiv:1905.06610 [hep-ph]].
%13 citations counted in INSPIRE as of 02 Oct 2022

%\cite{Bicudo:2020qhp}
\bibitem{Bicudo:2020qhp}
P.~Bicudo, N.~Cardoso, L.~Mueller and M.~Wagner,
%``Computation of the quarkonium and meson-meson composition of the $\Upsilon(nS)$  states and of the new $\Upsilon(10753)$ Belle resonance from lattice QCD static potentials,''
Phys. Rev. D \textbf{103}, no.7, 074507 (2021)
doi:10.1103/PhysRevD.103.074507
[arXiv:2008.05605 [hep-lat]].
%13 citations counted in INSPIRE as of 03 Oct 2022

%\cite{Ali:2019okl}
\bibitem{Ali:2019okl}
A.~Ali, L.~Maiani, A.~Y.~Parkhomenko and W.~Wang,
%``Interpretation of $Y_b (10753)$ as a tetraquark and its production mechanism,''
Phys. Lett. B \textbf{802}, 135217 (2020)
doi:10.1016/j.physletb.2020.135217
[arXiv:1910.07671 [hep-ph]].
%8 citations counted in INSPIRE as of 03 Oct 2022

%\cite{Yang:2009zzp}
\bibitem{Yang:2009zzp}
Y.~Yang, C.~Deng, J.~Ping and T.~Goldman,
%``S-wave Q Q anti-q anti-q state in the constituent quark model,''
Phys. Rev. D \textbf{80}, 114023 (2009)
doi:10.1103/PhysRevD.80.114023
%70 citations counted in INSPIRE as of 03 Oct 2022


%\cite{Vijande:2004he}
\bibitem{Vijande:2004he}
J.~Vijande, F.~Fernandez and A.~Valcarce,
%``Constituent quark model study of the meson spectra,''
J. Phys. G \textbf{31}, 481 (2005)
doi:10.1088/0954-3899/31/5/017
[arXiv:hep-ph/0411299 [hep-ph]].
%309 citations counted in INSPIRE as of 10 Aug 2022


%\cite{Vijande:2009pu}
\bibitem{Vijande:2009pu}
J.~Vijande and A.~Valcarce,
%``The rho-omega splitting in constituent quark models,''
Phys. Lett. B \textbf{677}, 36-38 (2009)
doi:10.1016/j.physletb.2009.05.017
[arXiv:0905.1817 [hep-ph]].
%5 citations counted in INSPIRE as of 03 Oct 2022



%\cite{Hu:2021nvs}
\bibitem{Hu:2021nvs}
X.~Hu and J.~Ping,
%``Investigation of hidden-charm pentaquarks with strangeness $S=-1$,''
Eur. Phys. J. C \textbf{82}, no.2, 118 (2022)
doi:10.1140/epjc/s10052-022-10047-z
[arXiv:2109.09972 [hep-ph]].
%7 citations counted in INSPIRE as of 10 Aug 2022

%\cite{Tan:2020ldi}
\bibitem{Tan:2020ldi}
Y.~Tan, W.~Lu and J.~Ping,
%``Systematics of $QQ{\bar{q}}{\bar{q}}$ in a chiral constituent quark model,''
Eur. Phys. J. Plus \textbf{135}, no.9, 716 (2020)
doi:10.1140/epjp/s13360-020-00741-w
[arXiv:2004.02106 [hep-ph]].
%31 citations counted in INSPIRE as of 10 Aug 2022

%\cite{Deng:2016ktl}
\bibitem{Deng:2016ktl}
W.~J.~Deng, H.~Liu, L.~C.~Gui and X.~H.~Zhong,
%``Spectrum and electromagnetic transitions of bottomonium,''
Phys. Rev. D \textbf{95}, no.7, 074002 (2017)
doi:10.1103/PhysRevD.95.074002
[arXiv:1607.04696 [hep-ph]].
%48 citations counted in INSPIRE as of 22 Oct 2022


%\cite{Hiyama:2003cu}
\bibitem{Hiyama:2003cu}
E.~Hiyama, Y.~Kino and M.~Kamimura,
%``Gaussian expansion method for few-body systems,''
Prog. Part. Nucl. Phys. \textbf{51}, 223-307 (2003)
doi:10.1016/S0146-6410(03)90015-9
%480 citations counted in INSPIRE as of 10 Aug 2022


\bibitem{Taylor:1970}
Howard S.~Taylor,
%``MODELS, INTERPRETATIONS, AND CALCULATIONS CONCERNING RESONANT ELECTRON SCATTERING PROCESSES IN ATOMS AND MOLECULES,''
Advan. Chem. Phys. \textbf{18}, 91-147(1970).
doi:10.1002/9780470143650.ch3
%480 citations counted in INSPIRE as of 10 Aug 2022

\bibitem{simons:1981}
J.~Simons,
J. Chem. Phys. \textbf{75}, 2465 (1981).
doi:10.1063/1.442271


%\cite{Hiyama:2018ukv}
\bibitem{Hiyama:2018ukv}
E.~Hiyama, A.~Hosaka, M.~Oka and J.~M.~Richard,
%``Quark model estimate of hidden-charm pentaquark resonances,''
Phys. Rev. C \textbf{98}, no.4, 045208 (2018)
doi:10.1103/PhysRevC.98.045208
[arXiv:1803.11369 [nucl-th]].


%\cite{Wu:2021ahn}
\bibitem{Wu:2021ahn}
Y.~Wu, X.~Jin, H.~Huang, J.~Ping and X.~Zhu,
%``Investigation of the meson-meson interaction,''
Phys. Rev. C \textbf{106}, no.2, 025204 (2022)
doi:10.1103/PhysRevC.106.025204
[arXiv:2109.14242 [hep-ph]].
%0 citations counted in INSPIRE as of 17 Oct 2022


%\cite{Hu:2022pae}
\bibitem{Hu:2022pae}
X.~Hu and J.~Ping,
%``Analysis of \ensuremath{\Omega}(2012) as a molecule in the chiral quark model,''
Phys. Rev. D \textbf{106}, no.5, 054028 (2022)
doi:10.1103/PhysRevD.106.054028
[arXiv:2207.05598 [hep-ph]].
%1 citations counted in INSPIRE as of 17 Oct 2022

%\cite{Chen:2021tnn}
\bibitem{Chen:2021tnn}
X.~Chen and Y.~Yang,
%``Doubly-heavy tetraquark states and *,''
Chin. Phys. C \textbf{46}, no.5, 054103 (2022)
doi:10.1088/1674-1137/ac4ee8
[arXiv:2109.02828 [hep-ph]].
%16 citations counted in INSPIRE as of 17 Oct 2022


\end{thebibliography}
\end{document}